\documentclass[fleqn,usenatbib]{mnras}

\usepackage{newtxtext,newtxmath}

\usepackage[T1]{fontenc}
\usepackage{ae,aecompl}
\usepackage{comment}

\usepackage{graphicx}	
\usepackage{amsmath}	
\usepackage{amssymb}	

\usepackage[dvipsnames]{xcolor}

\title[Beam Variations and Power Spectrum Estimation]{Calibration and 21-cm Power Spectrum Estimation in the Presence of Antenna Beam Variations}

\author[R.C. Joseph et al.]{
Ronniy C. Joseph,$^{1,2}$\thanks{E-mail: ronniy.joseph@icrar.org}
C. M. Trott,$^{1,2}$
R. B. Wayth$^{1,2}$
A. Nasirudin$^{1,2}$
\\
$^{1}$International Centre for Radio Astronomy Research - Curtin University, 1 Turner Avenue, Bentley WA 6102, Australia\\
$^{2}$ARC Centre of Excellence for All Sky Astrophysics in 3 Dimensions (ASTRO 3D), Perth, WA 6845\\
}

\date{Accepted 2019 November 28. Received 2019 November 28; in original form 2019 October 12}

\pubyear{2019}

\begin{document}
\label{firstpage}
\pagerange{\pageref{firstpage}--\pageref{lastpage}}
\maketitle

\begin{abstract}
Detecting a signal from the Epoch of Reionisation (EoR) requires an exquisite understanding of galactic and extra-galactic foregrounds, low frequency radio instruments, instrumental calibration, and data analysis pipelines. 
In this work we build upon existing work that aims to understand the impact of calibration errors on 21-cm power spectrum (PS) measurements. 
It is well established that calibration errors have the potential to inhibit EoR detections by introducing additional spectral features that mimic the structure of EoR signals. 
We present a straightforward way to estimate the impact of a wide variety of modelling residuals in EoR PS estimation. 
We apply this framework to the specific case of broken dipoles in Murchison Widefield Array (MWA) to understand its effect and estimate its impact on PS estimation.
Combining an estimate of the percentage of MWA tiles that have at least one broken dipole (15\%-40\%) with an analytic description of beam errors induced by such dipoles, we compute the residuals of the foregrounds after calibration and source subtraction. 
We find that that incorrect beam modelling introduces bias in the 2D-PS on the order of $\sim 10^3\, \mathrm{mK}^2 \,h^{-3}\, \mathrm{Mpc}^{3}$. Although this is three orders of magnitude lower than current lowest limits, it is two orders of magnitude higher than the expected signal. Determining the accuracy of both current beam models and direction dependent calibration pipelines is therefore crucial in our search for an EoR signal. 
\end{abstract}




\section{Introduction} 
\label{section:intro}
Detecting a redshifted neutral hydrogen signal from the Epoch of Reionisation (EoR) is one of the most promising probes into formation history of structure in the Universe. 
The signal enables us to directly observe the state of the intergalactic medium (IGM) over a wide range of cosmic time and indirectly study the sources that impact it. 
As the very first luminous sources light up the Universe, they heat up the IGM and subsequently reionise it. 
The redshifted 21-cm line gives us direct insight into the evolution of the IGM temperature and the morphology of the ionisation structures carved out by the first sources of light \citep{Morales2010, Pritchard2012, McQuinn2016, Furlanetto2016}. 

However, despite a large international effort by various telescope collaborations around the world, e.g the Murchison Widefield Array  \citep[MWA;][]{Tingay2013, Wayth2018}, the LOw Frequency ARray  \citep[LOFAR;][]{vanHaarlem2013} and the Donald C. Backer Precision Array for Probing the Epoch of Reionization \citep[PAPER; ][]{Parsons2010}, the signal has continued to elude a 21-cm power spectrum (PS) detection. 
The challenges faced by this experiment are large; foregrounds are 4-5 orders of magnitude brighter \citep{Jelic2008} and the instruments have a complex signal chain. Understanding the behaviour of the latest instruments is ongoing work, and continues to provide crucial input to our calibration strategies.
There already exists a large body of work on the residuals after direct subtraction of galactic and extra-galactic foregrounds, and their impact on the 21-cm PS. 
Most of this work treats the residuals of subtracted foregrounds as a source of {Gaussian} noise and studies how they affect the 21-cm PS assuming calibration leaves them unchanged  \citep{Liu2011, Trott2012, Dillon2013, Dillon2015, Trott2016, Murray2017}. 
However, it has been well studied that unmodelled foreground noise is non-Gaussian and that outliers in the tail-end of the noise distribution impact calibration on a non-negligible levels \citep{Kazemi2013, Ollier2017, Ollier2018}.
In particular, \citet{Barry2016} and \citet{Patil2016} show that sky based calibration in the presence of unmodelled foregrounds imparts additional spectral structure onto data, further inhibiting the detection of an EoR signal. \citet{Ewall_Wice2016} study this effect rigorously with a Gaussian approximation, and derive the imparted spectral structure on residual foregrounds due to modelling errors and found similar results.
Similarly for redundant calibration, we know that non-redundancies impart bias onto the calibration solutions. 
In \citet{Joseph2018a} we study this for position errors on calibration solutions, and \citet{Orosz2019} study the impact of positions errors and beam errors on the 21-cm PS and find that indeed non-redundancies also impart spectral structure that contaminates EoR detections. 
On top of that, redundant calibration needs external information set by sky based calibration to determine overall calibration parameters, hence the limitations of sky based calibration set a fundamental limit on the calibration accuracy of redundant calibration. 
This yet again introduces additional spectral structure \citep{Byrne2019}. 

In this paper we build on existing work to further study the impact of modelling errors on sky model calibration, and on 21-cm PS estimation. 
We derive a framework that enables rather simple propagation of errors into calibration solutions and the EoR power spectrum. We focus our attention on what is undoubtedly the next challenge in the EoR experiment: the performance of individual elements. 
We describe the errors introduced by broken dipoles in the MWA and make an informed estimate on the contamination we expect. 
In section \ref{section: calibration} we discuss calibration, source subtraction and signal estimation; in sections \ref{section: uncalibrated_covariance}, \ref{section: gain_covariance} and \ref{section: corrected_residuals} we derive the covariance matrices that describe our errors, propagate those forward to our gain solutions, and combine the two to derive the frequency structure of the calibrated residuals, respectively. 
In section \ref{section: results} we present results from the derived framework, we compare sky and beam modelling errors, compare the results to a fiducial EoR signal, and we estimate the impact of broken dipoles in the MWA EoR experiment. 
We discuss the applicability of this framework to other sources of error, and the implications of these results in an EoR context in section \ref{section: Discussion}.

Throughout this paper, our notation is as follows: lowercase bold letters $\mathbf{v}$ describe vectors, uppercase bold letters $\mathbf{C}$ describe matrices, $^{\dagger}$ denotes the Hermitian transpose, $^{*}$ denotes complex conjugation, and $i$ is the imaginary unit.

\section{EoR Signal Estimation with Gain Calibration Errors}
\label{section: calibration}
The aim of calibration is to mitigate all effects that inhibit us from estimating the true sky intensities $I(\mathbf{l}, \nu)$, where $\mathbf{l}$ is the sky coordinate vector and $\nu$ is the observing frequency. In the flat sky approximation we can relate the sky intensity $I(\mathbf{l}, \nu)$ to the complex visibilities $V(\mathbf{u}, \nu)$ measured by a pair of antennas in an interferometer with baseline separation $\mathbf{u}$ through
\begin{equation}
\begin{aligned}
V(\mathbf{u}, \nu) = \int g_{p} g^{*}_{q}\, b_{p}(\mathbf{l},\nu) b^{*}_{q} (\mathbf{l},\nu)\, I(\mathbf{l}, \nu)\, e^{-2\pi i\mathbf{u}\cdot \mathbf{l}}\, d^2\mathbf{l},
\end{aligned}
\label{eq: full_measurement_equation}
\end{equation}

where $g_{p}$ is the complex-valued gain of antenna $p$, and $b_{p}(\mathbf{l},\nu)$ is the corresponding beam voltage response. There are currently two popular methods of calibration; sky based calibration and redundant calibration. 

Sky based calibration uses a model of the beam and the sky intensity to predict the visibilities $V_{pq}$  measured by a pair of antennas $p$ and $q$, and uses these model visibilities to solve for the unknown antenna gains $g_{p}$ by minimising the squared differences ($L^2$-norm) between the model and the data

\begin{equation}
\begin{aligned}
\chi^2  = \sum_{pq} \lvert V^{\mathrm{data}}_{pq} - g_{p} g_{q}^* V^{\mathrm{model}}_{pq} \rvert^2.
\end{aligned}
\label{eq: sky_based_cal_minimisation}
\end{equation}

Redundant calibration relies on having multiple identical baselines in arrays with a regular lay-out.
These copies measure the same visibility, and therefore minimising the difference
between the visibilities by varying the antenna gains in such groups allows us to find both the unknown antenna gains $g_{p}$ and unknown redundant visibilities $V_{\alpha}^{\mathrm{true}}$ without the need for a sky model.

\begin{equation}
\begin{aligned}
\chi^2  = \sum_{\alpha} \sum_{pq,\alpha} \lvert V^{\mathrm{data}}_{pq} - g_{p} g_{q}^* V^{\mathrm{true}}_{\alpha} \rvert^2.
\end{aligned}
\label{eq: redundant_cal_minimisation}
\end{equation}

In general, these gains, or antenna responses, are direction dependent, i.e. they can capture deviations from the beam model or distortions by the ionosphere. A direction dependent calibration approach is, however, limited by the number of directions it can solve for. This limit is set by the number of bright sources available for calibration and computational costs. The number of directions ranges from 5-100 for current calibration pipelines. Throughout this work we focus on the direction independent gains as a simplification of the problem, noting that we can describe uncorrected directions by the same perturbative approach we are taking. The direction independent gains describe the global response of the antenna and signal chain. Redundant calibration is inherently unable to solve for direction dependent effects.

In general, sky models are incomplete, redundant arrays have position errors, and there are variations in the antenna response. We can write our measured complex visibilities $V$, in the most general way, as a sum of a model $m$, residuals $r$ that encompass errors on our model due to unmodelled sources, beam response variations, or even low level RFI \citep{Wilensky2019}, EoR signal $s$, and thermal noise $n$;

\begin{equation}
    \begin{aligned}
    V_{pq} = g_{p} g^{*}_{q} \left(m_{pq} + r_{pq} + s_{pq} \right) + n_{pq}.  
    \end{aligned}
\label{eq: data_model}    
\end{equation}

When we have data of the form in Equation~(\ref{eq: data_model}) and we calibrate using an incomplete model $m$, we inherently get incorrect gain estimates $\hat{g}_{p} = g_{p} + \delta g_{p}$ due to the presence of the residuals. 
When we apply these gain estimates to the data, see Equation~(\ref{eq: gain_corrected_data}), we get corrected visibilities $\hat{V}$ that contain corruptions that inhibit us from detecting the EoR signal \citep{Barry2016, Ewall_Wice2016, Patil2016}. 
In general, we can ignore the EoR signal at the calibration step, because it is several orders of magnitude weaker than the noise and the foreground residuals. 

\begin{equation}
    \begin{aligned}
    \hat{V}_{pq} = \frac{g_{p} g^{*}_{q}\left(m_{pq} + r_{pq} +s_{pq}\right) + n_{pq}}{\hat{g}_{p} \hat{g}^{*}_{q}}.  
    \end{aligned}
\label{eq: gain_corrected_data}    
\end{equation}

After correcting the data, we subtract the sky model. In practice, it is subtracted as part of an iterative calibration process, i.e. ``peeling'' \citep{Noordam2004}.
This leaves us with estimated data residuals $\hat{r}$ that contain, amongst others, the EoR signal

\begin{equation}
    \begin{aligned}
    \hat{r}_{pq} = \frac{g_p g^{*}_q\left(m_{pq} + r_{pq} + s_{pq}\right)+ n_{pq}}{\hat{g}_p \hat{g}^{*}_q} - m_{pq}.  
    \end{aligned}
\label{eq: calibrated_residuals}    
\end{equation}

From these residuals we estimate the 21-cm PS. However, Equation~(\ref{eq: calibrated_residuals}) contains more than the cosmological signal of interest and we will derive the covariance matrix of the additional residuals to understand their impact on our estimate of the 21-cm PS. We will study these errors from a power spectrum perspective and our approach is as follows:

\begin{enumerate}
    \item We first compute the data residual covariance matrix $\mathbf{C_{\mathrm{r}}}(u, \nu, \nu^{\prime})$ within a power spectrum bin $u$ over different frequencies. Specifically, we derive the contribution of beam errors due to broken dipoles (see section \ref{section: uncalibrated_covariance}). 
    \item We then use this to compute an approximation of the gain error covariance matrix $\mathbf{C_{\mathrm{g}}}$. Instead of computing the gain error per antenna, we compute averaged gain error covariance matrix for each power spectrum bin (see section \ref{section: gain_covariance}).
    \item Finally, we combine the two results to derive the covariance matrix of the gain-calibrated and source-subtracted residuals $\mathbf{C_{\mathrm{\hat{r}}}}$ (see section \ref{section: corrected_residuals}).
\end{enumerate}

 To estimate how each error contributes to a bias in the EoR power spectrum we propagate these covariances from frequency-space forward to PS space. 
A Fourier transform over frequency is a linear operation that can be described by a complex matrix $\mathcal{\mathbf{F}}$ applied to our complex data vector containing frequency data. Hence, the covariance of the Fourier transformed data can be computed using standard linear error propagation: $\mathcal{\mathbf{F}}^{\dagger} \mathbf{C}\mathcal{\mathbf{F}}$. 
The variance of this propagated covariance matrix describes the added power due to these errors. The off-diagonals describe how this power correlates between different Fourier modes.

\section{The Residual Covariance Matrix}
\label{section: uncalibrated_covariance}

In this section we derive the different contributions to the residual covariance matrix $\mathbf{C_{\mathrm{r}}}$.
To derive these matrices we start with the general covariance of visibilities,

\begin{equation}
\begin{aligned}
\mathbf{C}_{\mathrm{r}} = \mathrm{Cov}[V(\mathbf{u}, \nu), V(\mathbf{u'},\nu')].
\end{aligned}
\label{eq:general_covariance}
\end{equation}

To derive the residual covariance $\mathbf{C}_{\mathrm{r}}$ we assume we can separate this into a covariance between different baselines, and a covariance within a given baseline between different frequencies \citep{Liu2014}. In general, baselines with a separation $\lvert \mathbf{u}_{1} - \mathbf{u}_{2} \rvert$ larger than the size of the Fourier transform of the primary beam decorrelate, and it suffices to compute the frequency covariance alone per $u$-bin.

We consider three contributions to the residual covariance: the covariance due to the sky $\mathbf{C}_{\mathrm{sky}}$ that describes the error due to unmodelled sources, the noise covariance $\mathbf{C}_{\mathrm{n}}$ that describes the error due to thermal noise, and the beam covariance $\mathbf{C}_{\mathrm{beam}}$ that describes the error due to deviations from the ideal beam model 

\begin{equation}
\begin{aligned}
\mathbf{C}_{\mathrm{r}} = \mathbf{C}_{\mathrm{sky}} +\mathbf{C}_{\mathrm{beam}} +\mathbf{C}_{\mathrm{n}}. 
\label{eq: separated_data_covariance}
\end{aligned}
\end{equation}

The noise covariance is independent from all other terms, and its structure is well known. We will not discuss it further in this paper.

\subsection{Sky Covariance Matrix}
The sky covariance matrix $\mathbf{C}_{\mathrm{sky}}$ for a baseline at different frequencies has been well studied \citep{Liu2011, Dillon2013, Trott2016, Murray2017}. 
It describes the noise due to unmodelled sources and how this noise correlates between different frequency channels. 
We assume that an infinitesimal patch of sky $\mathrm{d}^2\mathbf{l}$ contains a number of sources drawn from a Poisson distribution $\tilde{N}\sim \mathrm{Poisson}(\mathrm{d}N/\mathrm{d}S dS \mathrm{d}^2\mathbf{l})$. 
The intensity of this patch is given by the first moment $\mu_1$ of the source count distribution

\begin{equation}
    \mu_n = \int_{0}^{S_{\mathrm{max}}} S^{n} \frac{\mathrm{d}\tilde{N}}{\mathrm{d}S}\mathrm{d}S. 
\end{equation}

We model the differential source counts $\mathrm{d}\tilde{N}/\mathrm{d}S$ with a broken power law model \citep{DiMatteo2002, Ewall_Wice2016, Murray2017} to match observations in different flux regimes at low frequencies \citep{Gervasi2008,Intema2011,Franzen2016, Williams2016},

\begin{equation}
\begin{aligned}
\frac{\mathrm{d}N}{\mathrm{d}S} =   
\begin{cases}
k_{1} S^{-\beta_{1}} & \text{if $ S_{\mathrm{low}}\leq S< S_{\mathrm{mid}}$} \\
k_{2} S^{-\beta_{2}} & \text{if $ S_{\mathrm{mid}}\leq S< S_{\mathrm{high}}$} \\
\end{cases}
.
\end{aligned}
\label{source_counts}
\end{equation}

With this broken power law model we want to capture the difference between the distribution of modelled and unmodelled sources. We use $k_1 = k_2 = 4100$, $\beta_1 = 1.59$, $\beta_2 = 2.5 $, $S_{\mathrm{low}} = 100\,  \mathrm{mJy}$, $S_{\mathrm{mid}} = 1\, \mathrm{Jy}$, and $S_{\mathrm{high}} = 10\, \mathrm{Jy}$. 
For this model, we assume all sources above $1\,\mathrm{Jy}$ are used in calibration and all sources below that threshold are unmodelled. This results in the following expression for the sky covariance matrix,

\begin{equation}
\begin{aligned}
	\mathbf{C}_{\mathrm{sky}} = 2\pi  (f_{0} f^{\prime}_{0})^{-\gamma} \mu_{2} \Sigma^2 \exp{(-2\pi^2 u^2 \Delta f^2 \Sigma^2)}
\end{aligned}
.
\label{eq: sky_covariance}
\end{equation}

Here, $\mu_2$ is the second moment of the source count distribution, $\gamma$ is the power law index that models the spectral energy distribution of each source, $f_{0} = \nu/\nu_0$ where $\nu_0$ is some reference frequency, e.g. the lowest frequency in the bandwidth, $\Delta f = f_{0} - f^{\prime}_{0} $. Following the notation of \citet{Murray2017}, $\Sigma$ contains products of the beam widths $\sigma$ at different frequencies

\begin{equation}
    \Sigma^2 = \frac{\sigma^{2}\sigma^{\prime 2}}{\sigma^{2} +\sigma^{\prime 2}}.
\end{equation}

\subsection{Beam Covariance Matrix}
\label{sec: beam_covariance_derivation}
Here, we derive the beam perturbation covariance matrix. 
Similarly to the derivation of the sky covariance, we start out by taking equation~(\ref{eq: full_measurement_equation}) under ideal gains $g= 1$, and assume we have a modelled sky intensity $I$ and an unmodelled component $\delta I$. 
We extend this by adding a perturbation $\delta b$ to the response of one antenna $b_{p}$, i.e. $b_{p} = b + \delta b_{p}$. 
We leave the other antenna responses as perfect,

\begin{equation}
\begin{aligned}
V(\mathbf{u}, \nu) = \int b \big( b + \delta b)^*(I + &\delta I) \,\times e^{-2\pi i\mathbf{u}\cdot \mathbf{l}}\, d^2\mathbf{l}. \\
\end{aligned}
\label{eq: perturbed_beam_measurement_equation_simplified}
\end{equation}

We have implicitly written the sky, the beam and their perturbations as functions of sky coordinate $\mathbf{l}$ and frequency $\nu$ for brevity. The extra source of noise $\delta V(\mathbf{u}, \nu)$ is the sum of unmodelled components in the visibility
\begin{equation}
\begin{aligned}
\delta V(\mathbf{u}, \nu) = \int \Big(  b \delta b^{*} I  +b b^{*} \delta I  + b \delta b^{*} \delta  I \Big) \, e^{-2\pi i\mathbf{u}\cdot \mathbf{l}}\, d^2\mathbf{l}. \\
\end{aligned}
\label{eq: visibility_perturbation}
\end{equation}

In this derivation, we assume the beam perturbation $\delta b$, the modelled sky intensity $I$, and unmodelled sky $\delta I$ are random variables. We rewrite equation~(\ref{eq:general_covariance}) by dividing the sky in voxels and write equation~(\ref{eq: visibility_perturbation})
as

\begin{equation}
\begin{aligned}
\delta V(\mathbf{u}, \nu) &= \sum_{p} \left( b_{p} \delta b^{*}_{p} I_{p} + b_{p} b^{*}_{p} \delta I_{p} + b_{p} \delta b^{*}_{p} \delta I_{p} \right) \, e^{-2\pi i\mathbf{u}\cdot \mathbf{l}_{p}}\, d^2\mathbf{l}_{p}. \\
\end{aligned}
\label{eq: perturbed_beam_measurement_equation_discrete}
\end{equation}

Combining equations~(\ref{eq:general_covariance}) and (\ref{eq: perturbed_beam_measurement_equation_discrete}), we rewrite the covariance as the sum of the covariances between the Fourier transforms of different sky voxels ${p}$ and ${q}$
\begin{equation}
    \begin{aligned}
    \mathbf{C}_{\mathrm{r}} &= \sum_{p} \sum_{q} \mathbf{C}_{{pq}}\\
    \end{aligned}
    .
    \label{eq: covariance_voxelised}
\end{equation}

where the covariance between different voxels is given by
\begin{equation}
    \begin{aligned}
    \mathbf{C}_{ij} &= \mathrm{Cov}[(b_{p} \delta b^{*}_{p} I_{p} + b_{p} b^{*}_{p} \delta I_{p} + b_{p} \delta b^{*}_{p} \delta I_{p}) \, e^{-2\pi i\mathbf{u}\cdot \mathbf{l}_{p}}\, d^2\mathbf{l}_{p},\\ 
    &\quad\quad\quad(b^{\prime}_{q} \delta b^{\prime*}_{q} I^{\prime}_{q} + b^{\prime}_{q} b^{\prime*}_{q} \delta I^{\prime}_{q} + b^{\prime}_{q} \delta b^{\prime*}_{p} \delta I^{\prime}_{q}) \, e^{-2\pi i\mathbf{u^{\prime}}\cdot \mathbf{l}_{q}}\, d^2\mathbf{l}_{q}]
    \end{aligned}
    .
\end{equation}

If we extract the constant terms from the covariance, we can focus on the stochastic terms, i.e. the modelled and unmodelled fluxes, and the beam perturbations.

\begin{equation}
    \begin{aligned}
    \mathbf{C}_{ij} = b_{p} b^{\prime *}_{q} e^{-2\pi i(\mathbf{u}\cdot \mathbf{l}_{p}  - \mathbf{u}^{\prime}\cdot \mathbf{l}_{q})} \times  \mathrm{Cov}&[\delta b^{*}_{p} I_{p} + b^{*}_{p} \delta I_{p} + \delta b^{*}_{p} \delta I_{p}, \, \\ &\delta b^{\prime *}_{p} I^{\prime}_{p} + b^{\prime *}_{p} \delta I^{\prime}_{p} + \delta b^{\prime*}_{p} \delta I^{\prime}_{p}]
    \end{aligned}
    .
\label{eq: covariance_stochastic_extracted}
\end{equation}

Using the formal definition of the covariance, $\mathrm{Cov}[X,Y] = \langle XY \rangle  - \langle X \rangle \langle Y \rangle$, and assuming $\delta b$, $I$, and $\delta I$ are independent we can expand this further and simplify (see appendix~\ref{apx: beam_covariance_matrix})

\begin{equation}
\begin{aligned}
\mathbf{C}_{ij} &= b_{p}  b^{\prime *}_{q} e^{-2\pi i(\mathbf{u}\cdot \mathbf{l}_{p}  - \mathbf{u}^{\prime}\cdot \mathbf{l}_{q})} \times 
\Big(
\langle \delta b^{*}_{p} \delta b^{'}_{q} \rangle \mathrm{Cov}[I_{p},I^{\prime}_{q}] \\
&+\left(b^{*}_{p} b^{\prime}_{q} + b^{*}_{p} \langle \delta b^{\prime *}_{q}\rangle + \langle \delta b^{*}_{p} \rangle b^{\prime *}_{q} + \langle \delta b^{*}_{p} \delta b^{'}_{q} \rangle \right)   \mathrm{Cov}[\delta I_{p}, \delta I^{\prime}_{q}] \\
&+\left(\langle I_{p} \rangle \langle I_{q}  \rangle + \langle I_{p}\rangle \langle \delta I^{\prime}_{q}\rangle  + \langle \delta I_{p} \rangle \langle I^{\prime}_{q} \rangle + \langle \delta I_{p} \rangle \langle  \delta I_{q} \rangle \right)\\
&\times \mathrm{Cov}[\delta b^{*}_{p},\, \delta b^{\prime *}_{q}] \Big ) 
\end{aligned}
.
\label{eq: covariance_worked_out}
\end{equation}

Using the source count distribution that describes the number of sources in a flux bin we can write $\mathrm{Cov}[I_{p}, I_{q}]$ as

\begin{equation}
\mathrm{Cov}[I_{p}, I_{q}] = (f^{}_{0} f^{\prime}_{0})^{-\gamma} \int S_{p} S_{q} \mathrm{Cov}[N_{p}, N_{q}] \mathrm{d}S
.
\end{equation}

 However, because different parts of the sky are independent realisations of a Poisson distribution, $\mathrm{Cov}[N_{p}, N_{q}]$ reduces to the mean for the same sky voxel ($\delta_{ij}\mathrm{d}N/\mathrm{d}S $). The 3rd term in Equation~(\ref{eq: covariance_worked_out}) then results in the unmodelled sky covariance $\mathbf{C}_{\mathrm{sky}}$, see Equation~(\ref{eq: sky_covariance}). The remaining terms will be grouped in the beam covariance matrix $\mathbf{C}_{\mathrm{beam}}$. 

Combining Equations~(\ref{eq: covariance_voxelised}) and (\ref{eq: perturbed_beam_measurement_equation_discrete}), and integrating this over the sky and all fluxes $S$ we get  

\begin{equation}
    \begin{aligned}
    &\mathbf{C}_{\mathrm{r}} = \mathbf{C}_{\mathrm{sky}} + \mathbf{C}_{\mathrm{beam}}  \\
        \end{aligned}
    \label{eq: full_covariance}
\end{equation}

where the full form of the beam covariance matrix is given by

\begin{equation}
    \begin{aligned}    
    \mathbf{C}_{\mathrm{beam}} &= \left ( f_{0} f_{0}^{\prime}\right)^{-\gamma}  \mu_{2, \mathrm{m}} \int  \,\big \langle \delta b^{*}(\mathbf{l}, \nu) \delta b(\mathbf{l}, \nu^{\prime}) \big \rangle \\
    &\qquad \qquad \qquad  \times b(\mathbf{l},\nu) b^{*}(\mathbf{l}, \nu^{\prime}) e^{-2\pi i(\mathbf{u}  - \mathbf{u}^{\prime})\cdot \mathbf{l}} \mathrm{d}^2\mathbf{l}\\
    &  + \left ( f_{0} f_{0}^{\prime}\right)^{-\gamma} \mu_{2, \mathrm{r}} \int (b^{*} \langle \delta b^{\prime *}\rangle + \langle \delta b^{*} \rangle b^{\prime *} + \langle \delta b^{*} \delta b^{'} \rangle) \\
    &\qquad \qquad \qquad  \times b(\mathbf{l}, \nu) b^{*}(\mathbf{l}, \nu^{\prime}) e^{-2\pi i(\mathbf{u}  - \mathbf{u}^{\prime})\cdot \mathbf{l}} \mathrm{d}^2\mathbf{l} \\
    &  +\left ( f_{0} f_{0}^{\prime}\right)^{-\gamma}(\mu_{1, \mathrm{m}} + \mu_{1, \mathrm{r}})^2 \iint \mathrm{Cov}[\delta b^{*}(\mathbf{l}, \nu), \, \delta b^{*}(\mathbf{l}^{\prime}, \nu^{\prime})] \\
    &\qquad \qquad \quad  \times b(\mathbf{l}, \nu) b^{*}(\mathbf{l}^{\prime}, \nu^{\prime}) e^{-2\pi i(\mathbf{u}\cdot \mathbf{l}  - \mathbf{u}^{\prime}\cdot \mathbf{l}^{\prime})}\mathrm{d}^2\mathbf{l}\mathrm{d}^2\mathbf{l}^{\prime}.
    \end{aligned}
    \label{eq: full_beam_covariance_integral}
\end{equation}

To summarise, the beam covariance has three contributions: 
\begin{enumerate}
    \item The first contribution comes from the modelled sources and effectively describes the residuals due to subtraction of those sources with an incorrect beam model.
    \item The second term describes how the noise from the unmodelled sources $\mathbf{C}_{\mathrm{sky}}$ is modified by the beam perturbations.
    \item The last term describes the added covariance due to correlations between different parts of the beam.
\end{enumerate}
The flux density of the sky is uncorrelated between different locations on the sky.
However, because changes in the beam are in general correlated up to some correlation length, this introduces additional noise set by the mean flux of the sky. 
This last term is only important on $\lvert u \rvert$-scales on the order of this correlation length, i.e the diameter of the antenna. 
In general, $\mathbf{C}_{\mathrm{beam}}$ is zero if the antenna response is ideal or more generally when the modelled response is equal to the actual response. The solution to the integrals in Equation~(\ref{eq: full_beam_covariance_integral}) depends strongly on the form of beam perturbation.

\section{The Phased Array Beam Model}
We now derive the beam covariance matrix $\mathbf{C}_{\mathrm{beam}}$ for the ``missing'' dipole case in an MWA tile. The MWA consists of 128 tiles, and each tile is a $4\times4$ array of dipoles on a ground screen. The detailed steps of the derivation can be found in Appendix \ref{apx: missing_dipoles}. Here, we will only discuss the important steps for brevity. To derive the covariance of visibilities due to beam perturbations, we start out with the formal description of the beam response of a phased array $b_{\mathrm{tile}}$ consisting of N dipoles on a ground screen,

\begin{equation}
\begin{aligned}
b_{\mathrm{tile}} = b_{\mathrm{dipole}} \times \sum_{n =0}^N w_n \exp\Big[-2\pi i \mathbf{x}_{n}\cdot \mathbf{l}/\lambda\Big] . 
\end{aligned}
\label{eq: phased_array_beam}
\end{equation}

In this description we assume that the individual element electric field responses $b_{\mathrm{dipole}}$ are identical, and we can then multiply this single element beam with the array factor to create a compound beam. 
In this array factor $w_{n}$ is the element weight. For the zenith pointings considered in this work, these weights range from 0 to 1, however, in general these weights are complex. $\mathbf{x}_{n}$ is the location of the $n$th element with respect to the center of the phased array. 
For simplicity, we approximate the full tile beam as a frequency dependent Gaussian, following

\begin{equation}
\begin{aligned}
b_{\mathrm{tile}} = \exp[-\lvert\mathbf{l}\rvert^{2}/2\sigma^2(\nu)].
\end{aligned}
\label{eq: gaussian_beam_approximation}
\end{equation}

We define the width of the voltage beam as $\sigma = \sqrt{2}\epsilon c/D \nu$ with $\epsilon = 0.42$. 
We re-scale an Airy disk to a Gaussian width using $\epsilon$ and assume an MWA tile diameter of $D = 4\, \mathrm{m}$. The factor $\sqrt{2}$ ensures that the square of the voltage beam is consistent with descriptions of the beam used in the literature.

\subsection{A Broken MWA Dipole}
\begin{figure}
    \centering
    \includegraphics[width=0.45\textwidth, trim = 80 0 80 0]{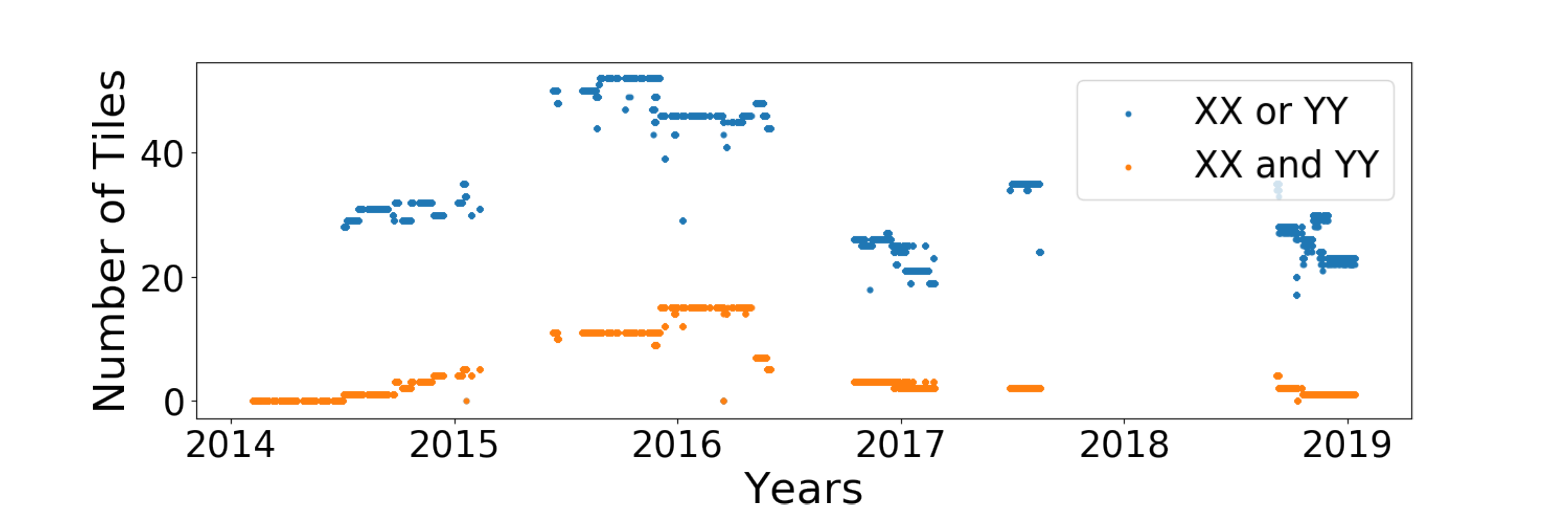}
    \caption{The number of MWA tiles with 1 broken dipole in a polarization and the number of tiles with 2 broken tiles in two different polarizations over several years of MWA EoR data. }
    \label{fig: broken_tiles_mwa}
\end{figure}

Although, there are many ways in which we can perturb the beam response, a common and relatively straightforward perturbation is the broken dipole case. 
Figure~(\ref{fig: broken_tiles_mwa}) shows the number of tiles with either one broken dipole in the X- or Y-polarisations, or two broken dipoles, one in each polarisation, for EoR observations over the past years. 
MWA tiles that have more than one broken dipole in the same polarization will be flagged and their data are not used.
At most 50 out of 128 tiles have been marked as having 1 broken dipole. 
Throughout this paper we choose a lower limit in this, adopting 25 broken dipoles which corresponds to $\sim30\%$ of the visibility data.

We describe the broken dipole perturbation to the beam $\delta b$ by subtracting the contribution of a missing dipole\footnote{Note: we explicitly consider phase offsets due to missing dipoles.}

\begin{equation}
\begin{aligned}
\delta b = - b_{\mathrm{dipole}} \times w_{\mathrm{broken}} \exp\Big[-2\pi i \mathbf{x}_{\mathrm{broken}} \cdot \mathbf{l}/\lambda\Big] . 
\end{aligned}
\label{eq: missing_dipole_beam}
\end{equation}

We approximate the response of the missing dipole towards the sky by a Gaussian, with a diameter $D$ that is 1/4 of the full tile, and a weight $w_{\mathrm{broken}} = 1$. 
This is equivalent to completely removing the dipole. To appropriately account for the contribution of a single dipole relative to the response of an $N$-element phased array, we normalise the dipole response by $N$:

\begin{equation}
\begin{aligned}
\delta b = \dfrac{1}{N} e^{-2\pi i \mathbf{x}_n \cdot l/\lambda } b_{\mathrm{dipole}}   
\end{aligned}.
\label{eq: missing_dipole_beam_approximation}
\end{equation}

We now have all the tools to derive the specific structure of $\mathbf{C}_{\mathrm{beam}}$ for the missing dipole case. 
We refer interested readers to Appendix \ref{apx: missing_dipoles} for details of this derivation, here we discuss results and implications for the 2D-PS. 

\begin{figure*}
    \centering
    \includegraphics[width = 1\textwidth]{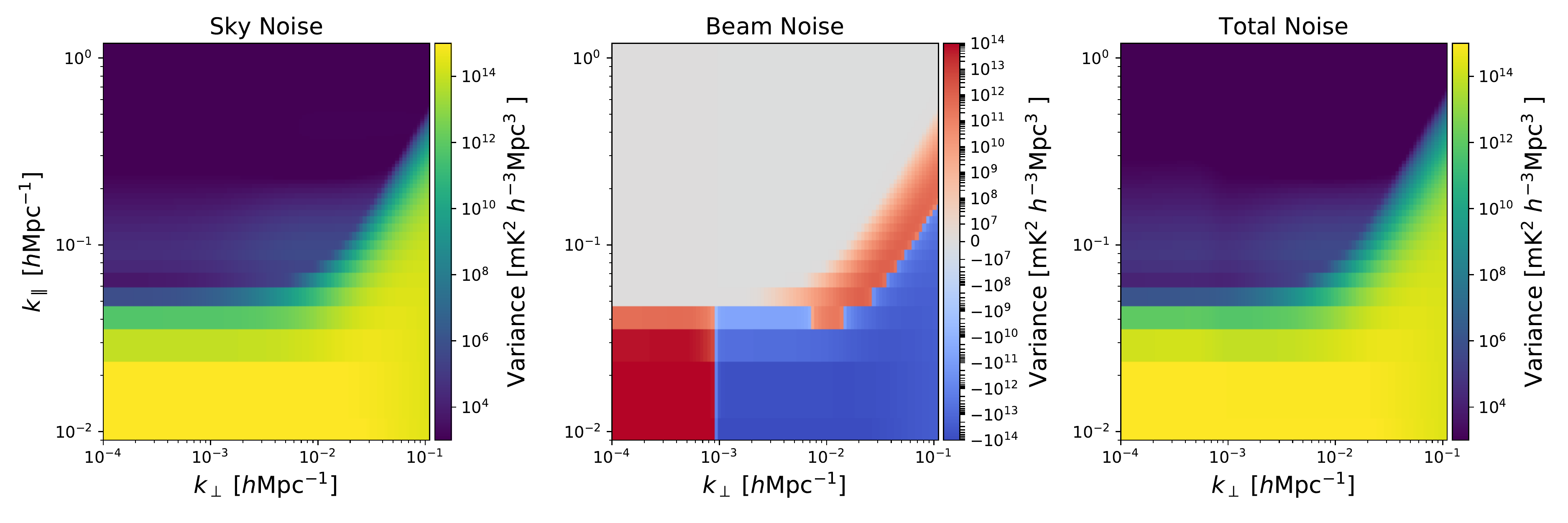}
    \caption{ The unmodelled sky covariance (left), the change in the covariance due to addition of beam perturbations to the covariance (middle), and the resulting total covariance (right). We show the 2D-PS for a 30 MHz bandwidth centred at 150 MHz with 251 frequency channels. Removing dipoles primarily takes away power from the wedge, because the array is less sensitive. However, this leaves us with modelled foreground residual power as those sources are incorrectly subtracted. This effect manifests itself primarily at the edge of the primary beam (or in the sidelobes of realistic phased arrays). This indicates contamination comes primarily from broken dipoles at the edge of a tile. The changing shape of the beam due to broken dipoles is a fairly large scale effect, both spatially and in frequency. Hence, it introduces power only at the the smallest $k_{\perp}$ and $k_{\parallel}$.}
    \label{fig: comparison_sky_and_beam_uncalibrated}
\end{figure*}

Figure~(\ref{fig: comparison_sky_and_beam_uncalibrated}) shows the 2D-PS for a 30 MHz bandwidth centred at 150 MHz with 251 frequency channels of the unmodelled sky variance, the beam variance and the total variance in cosmological units (see appendix \ref{apx: covariance_matrix_fourier_transform} for the conversion between frequency covariance matrices and the PS). 
To show all effects due to beam modelling errors, we have extended the range of $k_{\perp}$ beyond the conventional EoR range. 
The overall structure of the foreground wedge has not changed drastically, but if we look at the beam covariance component alone, we see that including beam modelling errors changes the noise in three regions. 
At small $k_{\perp}$, corresponding to baseline lengths shorter than the physical dimensions of an MWA tile, we see the contribution due to the correlations from different parts in the beam. 
We also see that there is overall less power in the wedge, because the missing dipoles decrease the sensitivity of the array as a whole, and we therefore see less of the foregrounds. 
Finally, we see the residuals of the modelled foregrounds on the edge of the foreground wedge. 
The largest relevant change due to missing dipoles is in this region, and is dominated by sources in the sidelobes. 
Note: beam errors do no intrinsically contaminate the EoR window, any visible excess is leakage due to the first sidelobe of the Fourier transform of the Blackman-Harris window (see appendix \ref{apx: covariance_matrix_fourier_transform}).   

\section{The Gain Error Covariance Matrix}
\label{section: gain_covariance}
Now that we have expressions for the data residuals, we explore how they propagate to the gain solutions during calibration.
We derive the covariance matrix of the averaged gain error $\delta g$ within a PS bin $u$. Sky model based calibration aims to solve

\begin{equation}
    \begin{aligned}
    \hat{g}_{p} \hat{g}^{*}_{q} m_{pq} = g_{p} g^{*}_{q} (m_{pq} + r_{pq}).
    \label{eq: calibration_equation}
    \end{aligned}
\end{equation}

We can rearrange this to solve\footnote{Note: to solve this system we need to split each entry into its imaginary and real component, and set a reference antenna. 
We compute all errors relative to that reference antenna, implying that the reference antenna should be error free or close to.} for the ratios of true and estimated gain solutions $\hat{g}_{p} = g_{p} + \delta g_{p}$ on the left hand side, and total signal over model visibilities on the right hand side  

\begin{equation}
    \begin{aligned}
    \frac{\hat{g}_{p} \hat{g}^{*}_{q}}{g_{p} g^{*}_{q}}  =  \frac{(m_{pq} + r_{pq})}{m_{pq}}.
    \label{eq: calibration_rearranged}
    \end{aligned}
\end{equation}

We assume ideal gains $g_{p} = 1$, and that the gain error is small $\delta g_{p} \ll g_{p} = 1$ for all $p$. Most of the signal is contained in the sky model, and the residuals are much smaller, and we can therefore expect small gain errors:

\begin{equation}
    \begin{aligned}
    \delta g_{p} + \delta g^{*}_{q}  = \frac{r_{pq}}{m_{pq}}
    \label{eq: calibration_linearised}
    \end{aligned}.
\end{equation}

Applying this to all baselines and corresponding antenna gain errors yields a system of equations that can be rewritten in matrix form $\mathbf{A} \mathbf{x} = \mathbf{y}$. 
The vector $\mathbf{y}$ contains the residual-to-model ratios $r_{pq}/m_{pq}$, and the vector $\mathbf{x}$ contains the gain errors $\delta g_{p}$. 
The array matrix $\mathbf{A}$ relates a baseline to the antennas that it is made up from. 
Its (pseudo-)inverse $\mathbf{A}^{-1}$ tells us how much each ratio $r_{pq}/m_{pq}$ in a baseline contributes to an error in a gain solution $\delta g_{p}$. 
It also implies that the gain error $\delta g_{p}$ is a weighted sum of the residual-to-signal ratio in each baseline,

\begin{equation}
    \begin{aligned}
    \delta g_{p} = \sum_n \frac{1}{w_{pn}}\frac{r_{n}}{m_{n}}.
    \end{aligned}
    \label{eq: gain_error_weighted}
\end{equation}

 Here, the weights $w_{in}$ are the entries of the inverse of the array matrix $\mathbf{A}^{-1}$, where the index $n$ runs over each baseline (instead of $ij$). 
 To compute the gain error covariance matrix $\mathbf{C_{g}}$, we need to compute $\mathrm{Cov}[\delta g, \delta g^{\prime}]$. 
 Noting that Equation~(\ref{eq: gain_error_weighted}) is a sum over different baselines, that in general fall in different $u$-bins, and assuming the covariance between different baselines is zero leaves us with

\begin{equation}
    \begin{aligned}
    \mathbf{C_{g}} = \sum_n \frac{1}{w^{2}_{n}} \mathrm{Cov} \left[ \frac{r_n}{m_n}, \frac{r^{\prime}_n}{m^{\prime}_n} \right].
    \label{eq: gain_error_covariance}
    \end{aligned}
\end{equation}

\begin{figure}
    \centering
    \includegraphics[width=0.4\textwidth]{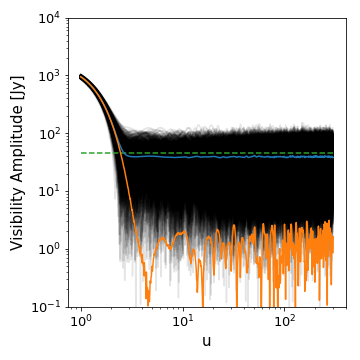}
    \caption{10 000 realisations of a stochastic sky and their amplitude across different $u$-scales.  The average amplitude of all realisation (blue), the amplitude of the average visibilities (orange), the amplitude of each individual realisation (black), and the expected sky r.m.s. for a power law distributed poisson sky (dashed green). On average all baseline do measure some signal, which is reasonably approximated the expected sky r.m.s.}
    \label{fig: mean_visibility_amplitudes}
\end{figure}

Instead of computing the covariance of a ratio of random variables, we approximate Equation~(\ref{eq: gain_error_covariance}) by replacing the model signal $m$  with the r.m.s. of the modelled sky $f_{0}^{-\gamma}\sqrt{\mu_{2,m}}$. 
This simplifies the expression for the gain error covariance matrix $\mathbf{C_{\mathrm{g}}}$. 
Figure~\ref{fig: mean_visibility_amplitudes} shows 10 000 visibility amplitude realisations of a stochastic sky, the amplitude of the mean visibility, the mean of the realised visibility amplitudes, and the sky r.m.s. From this we conclude that the sky r.m.s. provides a reasonable approximation to the modelled visibility $m_n$. We replace the modelled visibility $m_n$  with $f_{0}^{-\gamma}\sqrt{\mu_{2,m}}$, yielding    

\begin{equation}
    \begin{aligned}
    \mathbf{C_{g}} = \frac{(f_0 f^{\prime}_{0})^\gamma}{\mu_{2,m}} \sum_n \frac{1}{w^{2}_{n}} \mathrm{Cov} \left[r_n, r^{\prime}_n\right].
    \label{eq: gain_error_covariance_simplified}
    \end{aligned}
\end{equation}

 We study how data in one bin $u_{1}$ contributes to the error in a calibrated bin $u_{2}$; knowing that each baseline contributes to some gain solutions, each gain solution is applied to the data, and finally these data are binned into $\lvert u \rvert$-bins.  
 Instead of deriving what the gain error is on a single baseline, we derive how all uncalibrated PS bins change the structure of a calibrated bin through the ``averaged gain covariance''.

In our linearised approximation, each antenna gain error is a sum of $(N-1)$ baseline errors, i.e. all baselines in which that antenna participates. 
Each baseline therefore contributes $\sim 1/(N-1)$ to the gain solutions of the antennas it participates in. 
We smear this contribution out over all baselines, assuming each baseline contributes $\sim 1/N_b(N-1)$ to all gain solutions.  
Two of these antenna errors then propagate to a calibrated baseline. 
We compute the number of baselines in an uncalibrated bin that see baselines in a calibrated bin and relate that to the error weights $w(u)$, following        

\begin{equation}
    \begin{aligned}
    w(u) = \frac{1}{N-1}\frac{N_b(u)}{N_b(\mathrm{total})}. 
    \end{aligned}
\end{equation}

For an unrealistic array with perfect uniform $uv$-coverage, the number of baselines drops out, and the weights are $w = 1/N_{\mathrm{bins}}(N-1)$. For the MWA, we compute the actual distribution of baselines binned in the same way we bin all $k_{\perp}$.  

\begin{figure*}
    \centering
    \includegraphics[width=1\textwidth]{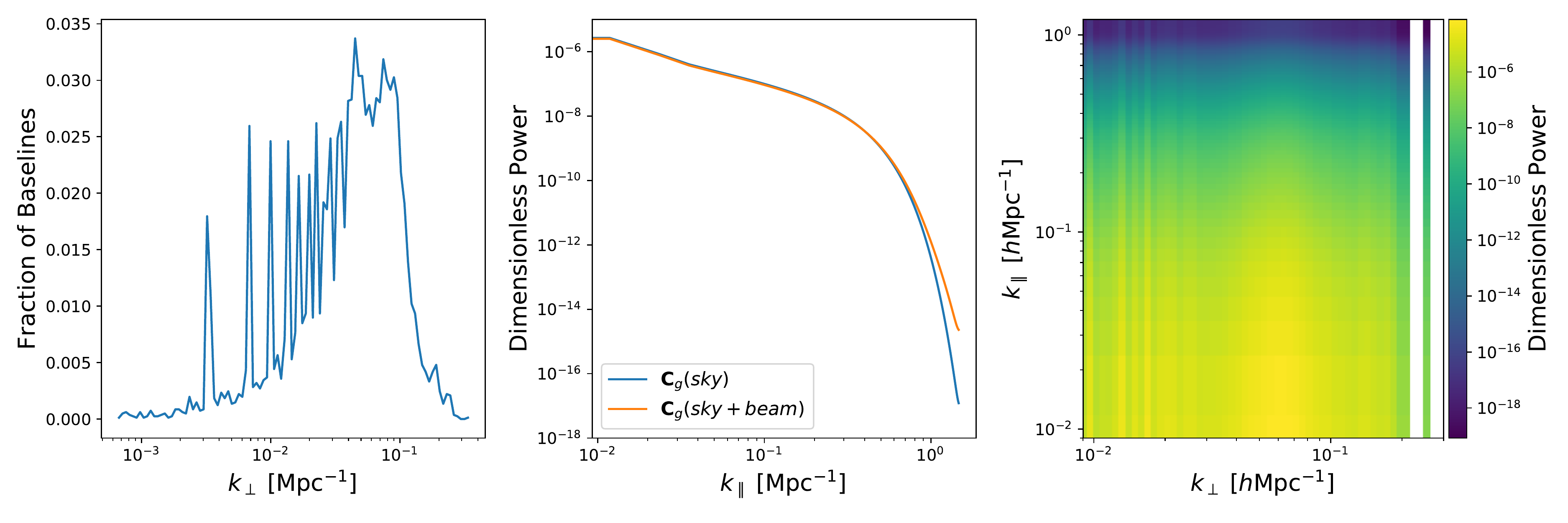}
    \caption{(left) A histogram of MWA Phase II compact baselines binned into power spectrum bins. The spikes arise from the redundant hexagons that produce many multiples of the same baselines. (middle) The Fourier transformed gain variance for an array with uniform $uv$-coverage with sky-only, and sky and beam errors. (right) The Fourier transformed gain variance for an MWA Phase II compact uv-coverage for sky model errors only. }
    \label{fig: gain_ps_convolution}
\end{figure*}

In Figure~(\ref{fig: gain_ps_convolution}), we show the binned MWA baselines, the variance of the Fourier transformed gain covariance assuming a uniform baseline distribution, and the Fourier transformed gain covariance for MWA Phase II Compact. For a uniform baseline distribution, the variance is the same for each $k_\perp$-bin; however, for the MWA the structure of the variance is different depending on how much it is coupled to other scales. In the next section, we demonstrate that this variance of the gain error effectively becomes a convolution window that smears out power from the foreground wedge into the EoR window. 

\section{The Gain Corrected Residual Covariance Matrix}
\label{section: corrected_residuals}
After obtaining our gain estimates $\hat{g}$, we apply them to the data and subtract our model visibilities $m$. 
This leaves us with our residual estimates $\hat{r}$ from which we aim to detect the EoR signal, see Equation~(\ref{eq: calibrated_residuals}). 
We assume the errors on our gains solutions are small $\delta g_{p}/g_{p} \ll 1$, enabling us to Taylor-expand ratios between our true gain solutions and the estimates $g_{p}/\hat{g}_{p}$. 
Grouping terms in products with either the model $m$ or the residuals $r$ results in

\begin{equation}
    \begin{aligned}
    \hat{r}_{pq} = -(\delta g_{p}  + \delta g^{*}_{q})m_{pq} + (1- \delta g_{p}  - \delta g^{*}_{q}) r_{pq}.
    \end{aligned}
    \label{eq: estimated_residuals_expanded}
\end{equation}

When we estimate the EoR signal, we grid and average these residual visibilities onto a $uv$-grid before Fourier transforming along the frequency-direction. We now want to compute the covariance of the averaged gridded residuals to understand the full impact of sky and beam modelling errors on the 21-cm PS 

\begin{equation}
    \begin{aligned}
    \mathbf{C_{\mathrm{\hat{r}}}}(u, \nu) &= \mathrm{Cov}[\mathbf{\hat{r}}, \mathbf{\hat{r}}^{\prime}] \\
    & = \langle \mathbf{\hat{r}} \mathbf{\hat{r}}^{\prime \dagger} \rangle - \langle\mathbf{\hat{r}}\rangle \langle \mathbf{\hat{r}}^{\prime} \rangle^{\dagger}.  
    \end{aligned}
    \label{eq: covariance_residuals}
\end{equation}

Equation~(\ref{eq: estimated_residuals_expanded}) keeps track off individual antenna $p$ and $q$ and how they impact a baseline. 
However, we are interested in the covariance of calibrated and model-subtracted data binned at scale $u$. 
The details of these final steps can be found in Appendix \ref{apx: linearised_gain_error_covariance}.
Here we only describe the general assumptions we made. 
To derive the covariance structure, we see the gains $g_{p}$ and $g_{q}$ as two realisations of a random variable within a $\lvert u \rvert$-bin, i.e. the two gain errors are independent. We also assume the gain error is independent from the model and data residuals of a certain $u$-bin. 
We justify the latter assumption because the gain error on a baseline or a $u$-bin is a linear combination of the residuals of different baselines. We already assumed that residuals in different $u$-bins are uncorrelated. This implies that a sum of residuals over different $u$-bins  decorrelates with the residuals of a single $u$-bin (if we include sufficiently independent residuals). We therefore assume that the gain error is independent from the residuals of the $u$-bin in which are computing the covariance of the calibrated residuals $\mathbf{C_{\mathrm{\hat{r}}}}$. When we combine Equation~(\ref{eq: estimated_residuals_expanded}) and (\ref{eq: covariance_residuals}), we can safely drop the expectation values of the estimated residuals $\langle \mathbf{\hat{r}} \rangle$, because they integrate to zero for baselines longer than the tile diameter. Under these assumptions, the covariance of calibrated data residuals reduces neatly to

\begin{equation}
    \begin{aligned}
    \mathbf{C_{\mathrm{\hat{r}}}} &= 2\mathbf{C}_{\mathbf{g}}\odot\mathbf{C}_{\mathbf{m}} + (1 + 2\mathbf{C}_{\mathbf{g}})\odot\mathbf{C}_{\mathbf{r}}.  
    \end{aligned}
    \label{eq: calibrated_residuals_final}
\end{equation}
This final equation compactly describes how calibration propagates errors on our data model into the data product from which we estimate an EoR signal.
A product of covariance matrices is a fairly unusual expression. In this specific case, $\mathbf{C_{g}}$ describes how much of the original covariances -- the model or residual covariance -- remain and how the correlation between different frequencies is changed. 
It is a product in PS space, and therefore becomes a convolution, between the  gain error and either the model covariance $\mathbf{C_{\mathrm{m}}}$ or $\mathbf{C_{\mathrm{r}}}$, in the PS and that smears out power from the foreground wedge throughout the PS. 
How much power is actually smeared out depends strongly on the errors that cause this correlation in the gain estimates.    

\section{Results}
\label{section: results}
\begin{figure}
    \centering
    \includegraphics[width=0.4\textwidth]{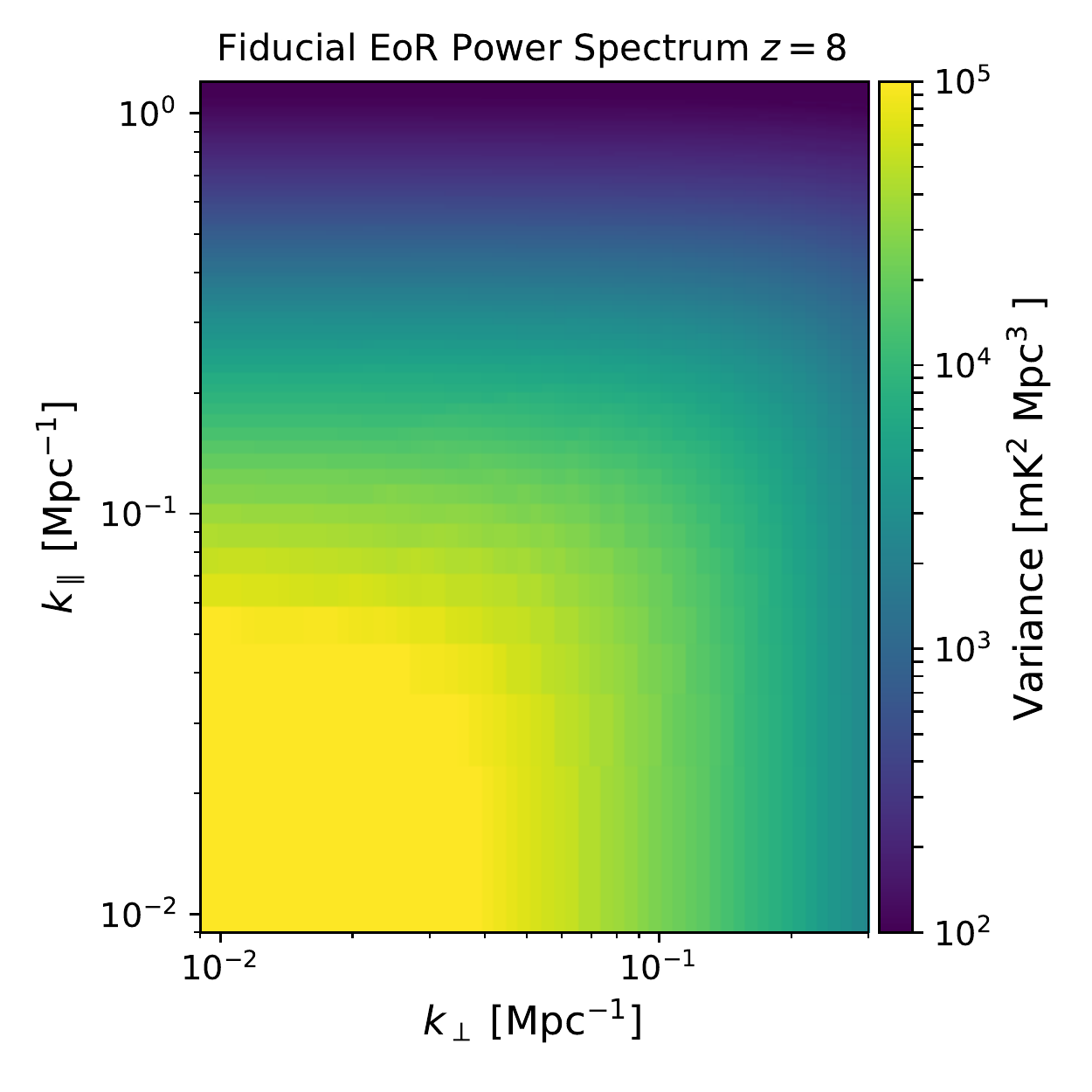}
    \caption{Fiducial EoR PS at redshift 8. We create this 2D PS by deprojecting a 1D EoR PS where the reionisation process is driven by faint galaxies \citep{Mesinger2016a}.  }
    \label{fig: fiducial_eor_power_specrum}
\end{figure}

Now that we have a relatively simple expression that describes the covariance of calibrated data residuals $\mathbf{C}_{\mathbf{\hat{r}}}$, we first apply this to an array with identical tile beams, i.e we consider unmodelled sky noise only, to compare with earlier studies on this problem. 
We then include beam errors to study how this extra source of modelling errors introduces contamination into the EoR window. 
We then compare these results to a fiducial EoR signal. 
We have taken a 1D-PS for a faint galaxy driven model at redshift $z\sim 8$ from \citet{Mesinger2016a}, and deprojected it into a 2D-PS assuming spherical symmetry, see Figure~(\ref{fig: fiducial_eor_power_specrum}). The code that generated the results presented here is publicly available, see \citet{joseph_software_2019}. 

\subsection{Sky Model Errors}
\begin{figure*}
    \centering
    \includegraphics[width=1\textwidth]{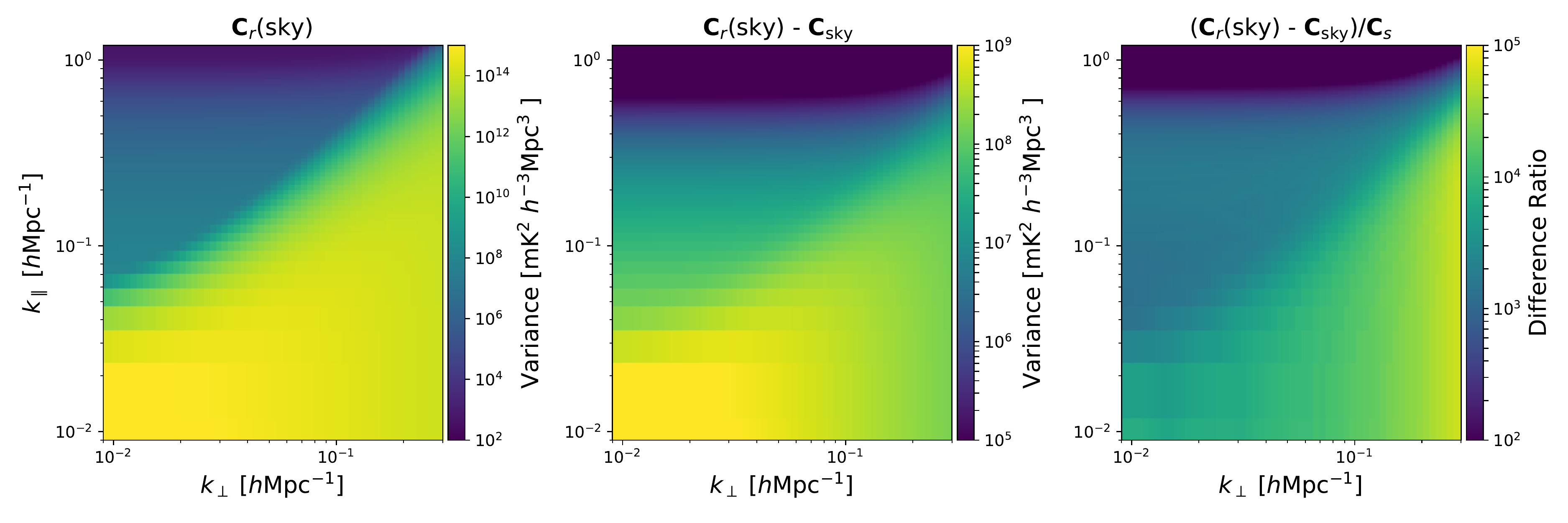}
    \caption{Comparing calibrated and uncalibrated residuals. Left: Calibrated and subtracted residuals with unmodelled sky noise only. Middle: Difference between calibrated and uncalibrated residuals. Right: Ratio between the difference  and a fiducial EoR PS.}
    \label{fig: comparison_sky_calibrated_uncalibrated}
\end{figure*}

Figure~(\ref{fig: comparison_sky_calibrated_uncalibrated}) shows the data residuals after calibration and model subtraction on the left, the difference between uncalibrated and calibrated residuals in the middle, and the ratio between that difference and our fiducial EoR signal on the right. 
The difference plot clearly shows the contamination into the EoR window introduced by the structure of the calibration errors. 
The additional bias we reproduce is similar to the results by \citet{Barry2016} and \citet{Ewall_Wice2016}, demonstrating that our formalism neatly reproduces earlier results. 
Here, we have only reproduced the results that describe the impact of unmodelled sources on a per frequency channel calibration strategy. 
We have not incorporated the mitigation strategies proposed to suppress excess noise due to these unmodelled sources. However, we note that the expected signal is strongest at the lowest $k$-modes hence mitigating leakage due to calibration is particularly important at these scales. 

\subsection{Comparing Sources of Error}
\begin{figure*}
    \centering
    \includegraphics[width=1\textwidth]{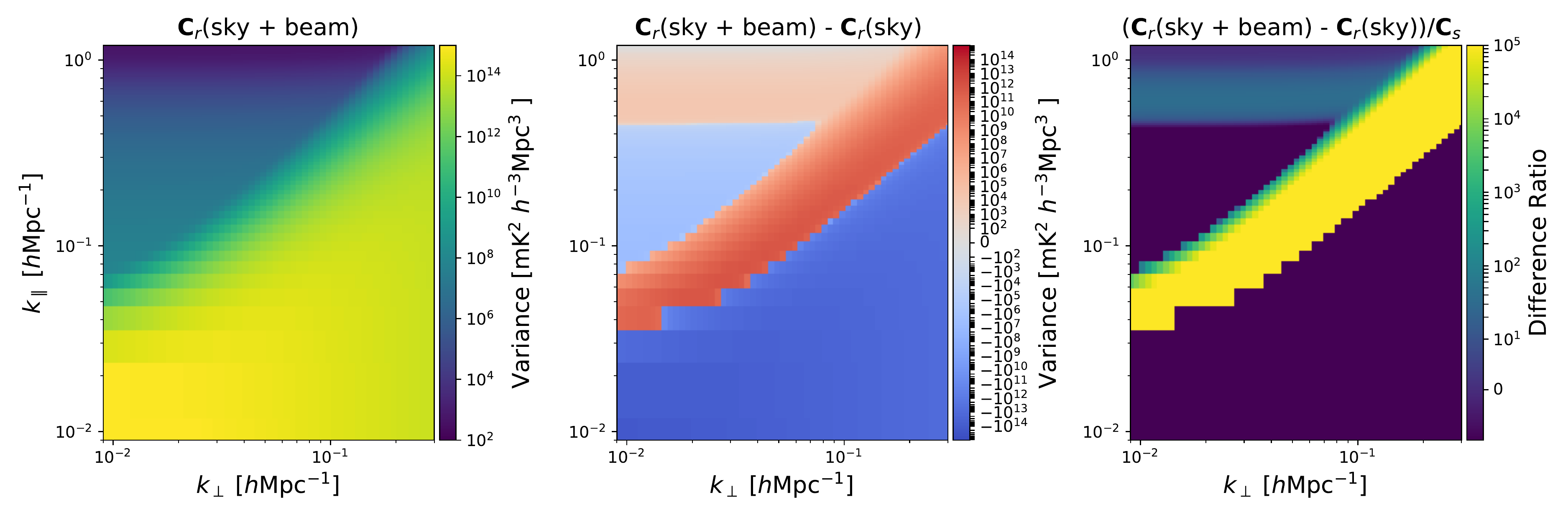}
    \caption{Comparing calibrated residuals with sky only and both sky and beam errors. Left: Calibrated Sky and Beam error residuals. Middle: Difference with Calibrated sky only residuals. Right: Ratio between difference and a fiducial EoR PS. Calibration in the presence of unmodelled sources adds significant contamination into the EoR window.}
    \label{fig: sky_and_beam_error_comparison_after_calibration}
\end{figure*}

We now apply this to data residuals including the beam covariance matrix $\mathbf{C}_{\mathrm{beam}}$. 
Figure~(\ref{fig: sky_and_beam_error_comparison_after_calibration}) shows the variance of the sum of calibrated residuals, the difference with  calibrated sky model only errors (the left most PS in Figure~(\ref{fig: comparison_sky_calibrated_uncalibrated})), and the ratio of this difference with a fiducial EoR signal. 
We see that the introduction of beam modelling errors adds power to the edge of the wedge as noted earlier in Figure~(\ref{fig: comparison_sky_and_beam_uncalibrated}). 
Now that we also include calibration, it smears out power further into the EoR window, causing a relative drop in power in the EoR window. 
This is not too surprising because Figure~(\ref{fig: gain_ps_convolution}) shows that beam errors changes the structure of the gain errors, such that they extend to larger $k_{\parallel}$. 

\begin{figure*}
    \centering
    \includegraphics[width=1\textwidth]{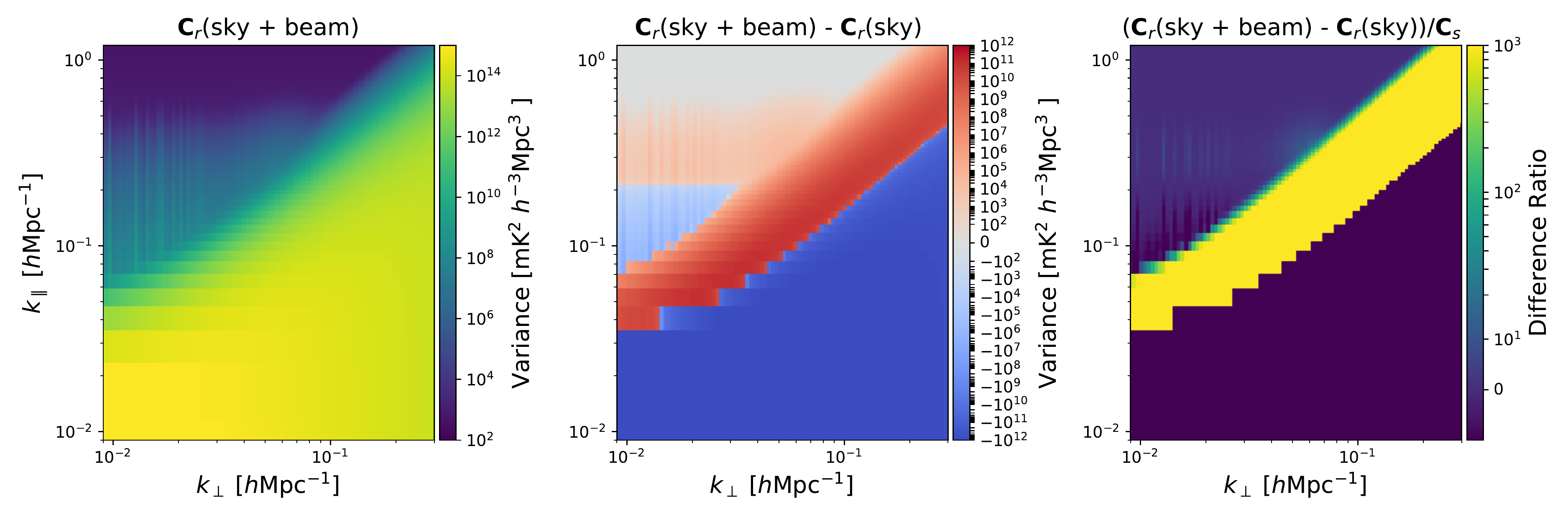}
    \caption{Expected MWA data contamination: Comparing calibrated residuals with sky only and both sky and beam errors. Left: Calibrated Sky and Beam error residuals. Middle: Difference with Calibrated sky only residuals. Right: Ratio between difference and a fiducial EoR PS.}
    \label{fig: MWA_sky_and_beam_error_comparison_after_calibration}
\end{figure*}

Ultimately, we want to estimate the expected contamination in the EoR window due to these errors under realistic circumstances. 
Figure~(\ref{fig: broken_tiles_mwa}) shows the number of broken MWA tiles during EoR observations, and from this we estimate $\sim30 \%$ of our visibility measurements to be contaminated. 
We appropriately down weight the beam covariance term $\mathbf{C}_{\mathrm{beam}}$ with a factor of $0.3^2$; one factor for both frequencies. 
We also use the baseline distribution in Figure~(\ref{fig: gain_ps_convolution}) to properly weight the covariance of each $u$-scale when we compute the gain error covariance matrix. Figure~(\ref{fig: MWA_sky_and_beam_error_comparison_after_calibration}), shows the expected results for an MWA-like data set. Down weighting the beam covariance changes the structure of the gain covariance in a way that the leakage does not extend that far into the EoR window. Nevertheless the additional contamination is still on the order of the expected EoR signal. 

\section{Discussion}
\label{section: Discussion}
The structure of the beam errors presented here is dominated by incorrectly subtracted sources towards the horizon. This hints that frequency structure mitigating techniques should rid us of most of the additional power due to beam modelling errors.
\citet{Barry2016} discussed these techniques after finding that incomplete sky models cause foreground power to be convolved with erroneous calibration solutions. They compute an expected level of contamination due to sky modelling error consistent with their simulation on the order of $\sim 10^7 \mathrm{mK^2\, h^{-3}\, Mpc^{-3}}$. We estimate contamination due to beam modelling errors to be 3 orders of magnitude lower than this. \citet{Patil2016} suggest multi-frequency calibration as a way around this. 
Multi-frequency calibration can enforce spectral smoothness on the solutions through a smoothness reguliser \citep{Yatawatta2015}, and subsequently decrease the variance reducing contamination of the EoR window. However, using all frequency information is computationally challenging and requires appropriate software architecture to overcome limited compute power \citep{Yatawatta2017}.

\citet{Ewall_Wice2016} derived the structure of the gain corrected residuals by keeping track of the baseline ordering. 
This leads them to directly relate the contamination from the longest baselines into the shorter baselines. 
When we compute the averaged gain covariance (Equation~\ref{eq: gain_error_weighted}), we take the average of the residuals covariances at different $k_{\perp}$ bins to describe the same effect. 
They suggest down weighting the longer baselines during calibration, as these baselines are the source of spectral contamination. 
On the other hand \citet{Patil2016} suggest excluding the shortest baselines for which we currently lack accurate models of the diffuse foregrounds. 
However, they also demonstrate that excluded baselines suffer from enhanced noise after calibration.
Creating diffuse sky maps similar to \citet{Eastwood2018} for the southern sky between $\nu \sim 35-70 \, \mathrm{MHz}$ is therefore crucial for accurate calibration. However, in future work we should also consider contamination from short baselines on which diffuse emission from the galaxy dominates due to the lack of such models.

This framework we derived can also be used to study the impact of non-redundancies on redundant calibration, i.e. the antenna position errors and beam variations. 
However, redundant calibration ultimately needs some external information to set overall gain parameters, e.g. absolute amplitudes and phase gradients see \cite{Wieringa1992, Liu2010} for more details. \citet{Byrne2019} show that incomplete sky models fundamentally limit the accuracy of redundant calibration solutions due to limitations in sky model based calibration. 
We expect this to be exacerbated by beam modelling errors that push power further into the window. \citet{Orosz2019} simulate redundant calibration for HERA including non-redundancies. 
They find that beam variations severely contaminate the EoR window. 
In future work, we will study what actually poses the largest hurdle to redundant arrays: inherent non-redundancies or incomplete sky models.

\cite{Li2018} study for the first time how redundant calibration and sky model calibration can be used optimally in MWA Phase II compact. 
They demonstrate that adding redundant calibration improves their results. In an effort to bridge the gap between redundant and sky based calibration, \citet{Sievers2017} developed the 'correlation calibration' framework that incorporates uncertainties on calibration models, e.g. sky model incompleteness, position offsets, and beam variations. 
There are currently tentative results that this is a very promising path forward; however, more work is required to properly compare this to current calibration techniques. 

Similar to earlier theoretical work we have not considered the non-Gaussian nature of the noise discussed by \citet{Kazemi2013, Ollier2017} and \citep{Ollier2018}. In this work we study the impact of the variance as the PS only sensitive to that. It is therefore not unreasonable to expect that the results presented here underestimate the level of contamination. 
We have also not considered mitigation strategies for EoR window contamination. 
In our estimation of added contamination due to beam modelling errors, we overestimate the expected errors for the MWA EoR experiment. 
Both MWA EoR pipelines RTS/CHIPS \citep{Mitchell2008, Trott2016} and FHD/$\epsilon$ppsilon \citep{Sullivan2012,Jacobs2016, Barry2019} incorporate direction dependent calibration. 
RTS/CHIPS uses information about broken dipoles explicitly to better model individual MWA tile beams. 
Similarly, many other calibration pipelines perform direction dependent calibration. 
However, we need to further quantify how much the responses vary from across an array similar to \citet{Line2018},
and how well direction dependent calibration captures variations in the beam to better estimate the expected contamination due beam modelling errors. Given the order of magnitude in which beam modelling errors manifests themselves, it seems very plausible that these errors are potentially our next systematic.

\section{Conclusion}
\label{section: conclusion}

Inspired by earlier theoretical work in this field, we have derived a relatively intuitive framework that neatly describes contamination in the EoR window due to calibration. 
We have reproduced earlier results and computed expected errors introduced by beam modelling errors. 
In this work, we have specifically focused on broken dipoles in the MWA as a perturbation to the model beam because this a relatively straightforward example. 
However, our results are applicable to a wide range of modelling errors, e.g. more complex beam variations and signal path variations, if analytic descriptions exists for these. 
We estimate that $\sim 15\%-40\%$ of the MWA tiles have at least one broken dipole. 
We have made a rough estimate on the order of magnitude $\sim 10^3\, \mathrm{mK}^2 \,h^{-3}\, \mathrm{Mpc}^{3} $ in which contamination by beam modelling errors manifests itself. 
These numbers are only indicative and we need to further quantify beam variations in-situ and determine how well current calibration techniques are able to account for this. However, we expect that these beam errors could potentially be our next limiting factor in the EoR experiment. 

\section*{Acknowledgements}
We thank the anonymous referee for constructive feedback that significantly improved this manuscript. RCJ thank Adrian Sutinjo for many linear algebra discussions, Andrew Williams, Marcin Sokolowski and Aman Chokshi for useful discussions on various aspects of broken MWA dipoles, and Torrance Hodgson for code optimisation suggestions.
This work was supported by the Centre for All Sky Astrophysics in 3 Dimensions (ASTRO 3D), an Australian Research Council
Centre of Excellence, funded by grant CE170100013. CMT is supported by an ARC Future Fellowship under grant FT180100321.
This research has made use of NASA's Astrophysics Data System.
We acknowledge the International Centre for Radio Astronomy
Research (ICRAR), a Joint Venture of Curtin University and
The University of Western Australia, funded by the Western
Australian State government. This scientific work makes use of
the Murchison Radio-astronomy Observatory, operated by
CSIRO. We acknowledge the Wajarri Yamatji people as the
traditional owners of the Observatory site. Support for the
operation of the MWA is provided by the Australian
Government (NCRIS), under a contract to Curtin University
administered by Astronomy Australia Limited.
This research has made use of NASA's Astrophysics Data System, Python \citep{Oliphant2007}, NumPy \citep{Oliphant2006, Walt2011}, and Matplotlib \citep{Hunter2007}.  




\bibliographystyle{mnras}
\bibliography{0catalogue}



\appendix

\section{Propagating Covariance Matrices to Power Spectrum space}
\label{apx: covariance_matrix_fourier_transform}
In this paper we derive covariance matrices that describe residuals in PS space. However, we are interested in the structures of residuals in PS space $(u, \eta)$. Normally, we apply a frequency taper $\Gamma$ to our data before Fourier Transforming our data. To compute the covariance in PS space we use the following transformation:

\begin{equation}
    \mathbf{\tilde{C}} = \mathcal{\mathbf{F}}^{\dagger} \Gamma(\nu)   \mathbf{C}\Gamma(\nu^{\prime})\mathcal{\mathbf{F}}
    .
\end{equation}

In this work we use a Blackmann-Harris function as a taper. Despite performing extremely well at surpressing sidelobes in PS space. It still has a non-neglible sidelobe within the EoR window, see Figure \ref{fig: comparison_sky_and_beam_uncalibrated}. This window function is also used in current MWA EoR pipelines. 

\section{The Beam Covariance Matrix}
\label{apx: beam_covariance_matrix}
Starting from equation~(\ref{eq: covariance_stochastic_extracted}) we can write out the cross-terms in the covariance matrix $\mathbf{C}_{ij}$.

\begin{equation}
    \begin{aligned}
\mathbf{C}_{ij} = b_{p} b^{\prime *}_{q}& e^{-2\pi i(\mathbf{u}\cdot \mathbf{l}_{p}  - \mathbf{u}^{\prime}\cdot \mathbf{l}_{q})} \times 
\Big(  
b^{*}_{p} b^{\prime}_{q} \mathrm{Cov}[\delta I_{p}, \delta I^{\prime}_{q}]  \\
& + 
b^{*}_{p} \mathrm{Cov}[\delta I_{p}, \delta b^{\prime *}_{q} I^{\prime}_{q}] 
+ 
b^{*}_{p} \mathrm{Cov}[\delta I_{p}, \delta b^{\prime *}_{q}  \delta I^{\prime}_{q}]\\
&+ 
b^{\prime}_{q} \mathrm{Cov}[\delta b^{*}_{p} I_{p}, \delta I^{\prime}_{q}] 
+ 
\mathrm{Cov}[\delta b^{*}_{p} I_{p}, \delta b^{\prime *}_{q} I^{\prime}_{q}] \\
&+
\mathrm{Cov}[\delta b^{*}_{p} I_{p}, \delta b^{\prime *}_{q} \delta I^{\prime}_{q}] 
+ 
b^{\prime}_{q}\mathrm{Cov}[\delta b^{*}_{p} \delta I_{p}, \delta I^{\prime}_{q}] \\
&+
\mathrm{Cov}[\delta b^{*}_{p} \delta I_{p}, \delta b^{\prime *}_{q} I^{\prime}_{q}] 
+ 
\mathrm{Cov}[\delta b^{*}_{p} \delta I_{p}, \delta b^{\prime *}_{q} \delta I^{\prime}_{q}] 
\Big ) \\
    \end{aligned}
\label{eq: appendix_covariance_fully_written}
\end{equation}

Using the formal definition of the covariance and the properties of the modelled component of the stochastic sky $I$, the unmodelled component of the sky $\delta I$, and our beam perturbations $\delta b$, i.e. the the sources that contribute to the intensity about the noise level are independent of the unmodelled sources. So we can further write out equation~(\ref{eq: appendix_covariance_fully_written}).

The first term results into the unmodelled sky covariance matrix $\mathbf{C}_{\mathrm{sky}}$ in equation~(\ref{eq: sky_covariance}). The second term becomes zero, because we can separate the averages over the modelled sky, the unmodelled sky, and the beam perturbation, that then cancel each other out.

\begin{equation}
\begin{aligned}
b^{*}_{p} \mathrm{Cov}[\delta I_{p}, \delta b^{\prime *}_{q} I^{\prime}_{q}] & = 
b^{*}_{p} \Big (\langle \delta I_{p} \delta b^{\prime *}_{q} I^{\prime}_{q}\rangle - \langle \delta I_{p} \rangle \langle \delta b^{\prime *}_{q} I^{\prime}_{q}\rangle \Big) \\
& = b^{*}_{p} \Big (\langle \delta I_{p} \rangle \langle \delta b^{\prime *}_{q}\rangle \langle I^{\prime}_{q}\rangle - \langle \delta I_{p} \rangle \langle \delta b^{\prime *}_{q} \rangle \langle I^{\prime}_{q}\rangle \Big)\\ 
& = 0
\end{aligned}
\end{equation}

The third term is describes a contribution to the modified sky noise, because the shape of the beam has been changed by a perturbation

\begin{equation}
\begin{aligned}
b^{*}_{p} \mathrm{Cov}[\delta I_{p}, \delta b^{\prime *}_{q} \delta I^{\prime}_{q}] 
& = b^{*}_{p} \Big( \langle \delta I_{p} \delta b^{\prime *}_{q} \delta I^{\prime}_{q}\rangle - \langle \delta I_{p} \rangle \langle \delta b^{\prime *}_{q} \delta I^{\prime}_{q}\rangle \Big) \\
& = b^{*}_{p}  \langle \delta b^{\prime *}_{q} \rangle  \mathrm{Cov}[\delta I_{p}, \delta I^{\prime}_{q}]  \\
\end{aligned}
\end{equation}

The fourth term becomes zero similarly to the second term.

\begin{equation}
\begin{aligned}
b^{\prime}_{q} \mathrm{Cov}[\delta b^{*}_{p} I_{p}, \delta I^{\prime}_{q}] &= 0
\end{aligned}
\end{equation}

The fifth term describes the covariance between different parts of the beam, i.e. a beam perturbation in generally changes large portions of the beam, this couples different parts of the sky.

\begin{equation}
\begin{aligned}
\mathrm{Cov}[\delta b^{*}_{p} I_{p}, \delta b^{\prime *}_{q} I^{\prime}_{q}] &=  \langle \delta b^{*}_{p} \delta b^{'}_{q} \rangle \mathrm{Cov}[I_{p},I^{\prime}_{p}] + \langle I_{p} \rangle \langle I_{q} \rangle \mathrm{Cov}[\delta b^{*}_{p},\, \delta b^{\prime *}_{q}]
\end{aligned}
\end{equation}

The sixth term describes can also be rewritten as the covariance between different parts of the beam
\begin{equation}
\begin{aligned}
\mathrm{Cov}[\delta b^{*}_{p} I_{p}, \delta b^{\prime *}_{q} \delta I^{\prime}_{q}] &=  \langle I_{p}\rangle \langle \delta I^{\prime}_{q}\rangle \mathrm{Cov}[\delta b^{*}_{p},\, \delta b^{\prime *}_{q}]
\end{aligned}
\end{equation}

The seventh term describes the modification to the sky noise due the perturbation of first frequency

\begin{equation}
\begin{aligned}
b^{\prime}_{q}\mathrm{Cov}[\delta b^{*}_{p} \delta I_{p}, \delta I^{\prime}_{q}] & = \langle \delta b^{*}_{p} \rangle b^{\prime *}_{q}   \mathrm{Cov}[\delta I_{p}, \delta I^{\prime}_{q}] 
\end{aligned}
\end{equation}

The eight' term describes the added covariance due the beam perturbations, similarly to the sixth term.

\begin{equation}
\begin{aligned}
\mathrm{Cov}[\delta b^{*}_{p} \delta I_{p}, \delta b^{\prime *}_{q} I^{\prime}_{q}] & = 
\langle \delta I_{p} \rangle \langle I^{\prime}_{q} \rangle \mathrm{Cov}[\delta b^{*}_{p},\, \delta b^{\prime *}_{q}]
\end{aligned}
\end{equation}

And finally the last term, similarly to the fift term, describes the added covariance due to different parts of the beam and how it couples different parts of the unmodelled sky together.

\begin{equation}
\begin{aligned}
\mathrm{Cov}[\delta b^{*}_{p} \delta I_{p}, \delta b^{\prime *}_{q} I^{\prime}_{q}] & = 
\langle \delta b^{*}_{p} \delta b^{'}_{q} \rangle \mathrm{Cov}[\delta I_{p}, \delta I^{\prime}_{q}]\\
&\qquad+ \langle \delta I_{p} \rangle \langle  \delta I^{\prime}_{q} \rangle \mathrm{Cov}[\delta b^{*}_{p},\, \delta b^{\prime *}_{q}]
\end{aligned}
\end{equation}

This leaves us with 6 additional terms on top of the well understood unmodelled sky noise.

\begin{equation}
\begin{aligned}
\mathbf{C}_{ij} &= b_{p}  b^{\prime *}_{q} e^{-2\pi i(\mathbf{u}\cdot \mathbf{l}_{p}  - \mathbf{u}^{\prime}\cdot \mathbf{l}_{q})} \times \Big(\langle \delta b^{*}_{p} \delta b^{'}_{q} \rangle \mathrm{Cov}[I_{p},I^{\prime}_{q}] \\
&+(b^{*}_{p} b^{\prime}_{q} + b^{*}_{p} \langle \delta b^{\prime *}_{q}\rangle + \langle \delta b^{*}_{p} \rangle b^{\prime *}_{q} + \langle \delta b^{*}_{p} \delta b^{'}_{q} \rangle)   \mathrm{Cov}[\delta I_{p}, \delta I^{\prime}_{q}] \\
&+(\langle I_{p} \rangle \langle I^{\prime}_{q}  \rangle + \langle I_{p}\rangle \langle \delta I^{\prime}_{q}\rangle  + \langle \delta I_{p} \rangle \langle I^{\prime}_{q} \rangle + \langle \delta I_{p} \rangle \langle  \delta I^{\prime}_{q} \rangle )\mathrm{Cov}[\delta b^{*}_{p},\, \delta b^{\prime *}_{q}] \Big ) 
\end{aligned}
\label{eq: total_data_covariance_voxel}
\end{equation}

\section{Beam Variations due to Missing Dipoles}
\label{apx: missing_dipoles}
First we need to compute the averaged beam perturbation $\langle \delta b(\mathbf{l}, \nu) \rangle $ and the averaged product of perturbations $\langle \delta b(\mathbf{l}, \nu)b(\mathbf{l}', \nu') \rangle $. 
We can calculate the averaged beam perturbation by simply averaging over the 16 different beam perturbations, in equation~(\ref{eq: missing_dipole_beam_approximation}). 
Although it is possible to write a single expression for these averages using the geometry of a square MWA tile, however, when we Fourier transform this we will end up with a sum over the different broken dipole realisations. 
Hence, we keep this simple formulation that describes a general tile lay-out.

\begin{equation}
\begin{aligned}
\langle \delta b(\mathbf{l}, \nu) \rangle = -\frac{1}{N^2}b_{\mathrm{d}} \sum_{n=1}^N e^{-2\pi i \mathbf{x}_{\mathrm{n}}\cdot \mathbf{l}/\lambda }
\end{aligned}
\label{eq: averaged_dipole_beam_perturbation}
\end{equation}

We can calculate the averaged product of beam perturbations at different frequencies in a similar fashion.

\begin{equation}
    \begin{aligned}
    \langle \delta b(\mathbf{l}, \nu) \delta b^*(\mathbf{l}', \nu') \rangle &= \frac{1}{N^3}b_{\mathrm{d}}(\mathbf{l}, \nu) b^*_{\mathrm{d}}(\mathbf{l}', \nu') \sum_{n=1}^N e^{2\pi i \mathbf{x}_{\mathrm{n}}\cdot \mathbf{l}/\lambda } e^{2\pi i \mathbf{x}_{\mathrm{n}}\cdot \mathbf{l}'/\lambda' } \\
    &= \frac{1}{N^3}b_{\mathrm{d}}(\mathbf{l}, \nu) b^*_{\mathrm{d}}(\mathbf{l}', \nu')  \sum_{n=1}^N e^{-2\pi i \mathbf{x}_{\mathrm{n}}\cdot( \mathbf{l}/\lambda - \mathbf{l}'/\lambda') }
    \end{aligned}
    \label{eq: averaged_product_dipole_beam_perturbation}
\end{equation}

Combining equations~(\ref{eq: averaged_dipole_beam_perturbation}), (\ref{eq: averaged_product_dipole_beam_perturbation}), and (\ref{eq: full_beam_covariance_integral}), we get:

\begin{equation}
    \begin{aligned}
    \mathbf{C}_{\mathrm{beam}} &= \mathbf{C}_{\mathrm{A}} - \mathbf{C}_{\mathrm{B}} - \mathbf{C}_{\mathrm{C}} + \mathbf{C}_{\mathrm{D}} - \mathbf{C}_{\mathrm{E}} \\
    \end{aligned}
    \label{eq: full_beam_covariance_integral_MWA}
\end{equation}

where all individual covariance components are given by

\begin{equation}
    \begin{aligned}
    &\mathbf{C}_{A} = 
     (\mu_{2, \mathrm{m}} + \mu_{2, \mathrm{r}})\left ( f_{0} f_{0}^{\prime}\right)^{-\gamma} \int  \,\big \langle \delta b^{*}(\mathbf{l}, \nu) \delta b(\mathbf{l}, \nu^{\prime}) \big \rangle b(\mathbf{l},\nu) b^{*}(\mathbf{l}, \nu^{\prime}) \\
     &\qquad \qquad \qquad \qquad\qquad \qquad \qquad\qquad  \times e^{-2\pi i(\mathbf{u}  - \mathbf{u}^{\prime})\cdot \mathbf{l}} \mathrm{d}^2\mathbf{l} \\
    &\mathbf{C}_{B} = \mu_{2, \mathrm{r}} \left ( f_{0} f_{0}^{\prime}\right)^{-\gamma}\int b^{*}(\mathbf{l}, \nu) \langle \delta b^{\prime *}(\mathbf{l}, \nu^{\prime})\rangle b(\mathbf{l}, \nu) b^{*}(\mathbf{l}, \nu^{\prime})\\
    & \qquad \qquad \qquad \qquad\qquad \qquad \qquad\qquad  \times e^{-2\pi i(\mathbf{u}  - \mathbf{u}^{\prime})\cdot \mathbf{l}} \mathrm{d}^2\mathbf{l} \\ 
    &\mathbf{C}_{C} = \mu_{2, \mathrm{r}}\left ( f_{0} f_{0}^{\prime}\right)^{-\gamma} \int \langle \delta b^{*}(\mathbf{l}, \nu) \rangle b^{\prime *}(\mathbf{l}, \nu^{\prime}) b(\mathbf{l}, \nu) b^{*}(\mathbf{l}, \nu^{\prime})\\
    &\qquad \qquad \qquad \qquad\qquad \qquad \qquad\qquad  \times e^{-2\pi i(\mathbf{u}  - \mathbf{u}^{\prime})\cdot \mathbf{l}} \mathrm{d}^2\mathbf{l}  \\
    &\mathbf{C}_{D} = (\mu_{1, \mathrm{m}} + \mu_{1, \mathrm{r}})^2\left ( f_{0} f_{0}^{\prime}\right)^{-\gamma} \iint \langle \delta b^{*}(\mathbf{l}, \nu) \delta b(\mathbf{l}, \nu^{\prime}) \big \rangle \\
    &\qquad \qquad \qquad \qquad \qquad \times b(\mathbf{l}, \nu) b^{*}(\mathbf{l}^{\prime}, \nu^{\prime}) e^{-2\pi i(\mathbf{u}\cdot \mathbf{l}  - \mathbf{u}^{\prime}\cdot \mathbf{l}^{\prime})}\mathrm{d}^2\mathbf{l}\mathrm{d}^2\mathbf{l}^{\prime} \\
    &\mathbf{C}_{E} = (\mu_{1, \mathrm{m}} + \mu_{1, \mathrm{r}})^2 \left ( f_{0} f_{0}^{\prime}\right)^{-\gamma}\iint \langle \delta b^{*}(\mathbf{l}, \nu) \rangle \langle \delta b(\mathbf{l}, \nu^{\prime}) \big \rangle \\
    &\qquad \qquad \qquad \qquad \qquad \times b(\mathbf{l}, \nu) b^{*}(\mathbf{l}^{\prime}, \nu^{\prime}) e^{-2\pi i(\mathbf{u}\cdot \mathbf{l}  - \mathbf{u}^{\prime}\cdot \mathbf{l}^{\prime})}\mathrm{d}^2\mathbf{l}\mathrm{d}^2\mathbf{l}^{\prime}. \\
    \end{aligned}
    \label{eq: beam_covariance_split}
\end{equation}

This leaves us with 6 integrals to solve. We combine the covariance terms $A - D$ in equation~(\ref{eq: beam_covariance_split}) with (\ref{eq: averaged_dipole_beam_perturbation} ) and (\ref{eq: averaged_product_dipole_beam_perturbation}), and write the product of beams, into one single Gaussian. This enables us to use the Hankel transform for each broken dipole realisation. 

\begin{equation}
    \begin{split}
    \mathbf{C}_{\mathrm{A}} &= \frac{2\pi \Sigma_A^2 (\mu_{2,m} + \mu_{2,r}) \left ( f_{0} f_{0}^{\prime}\right)^{-\gamma}}{N^3 } \times \\
    &\qquad\sum_{n=0}^{N} \exp{\big(-2 \pi^2\Sigma_A^2 \lvert \mathbf{u} - \mathbf{u}^{\prime} + \mathbf{x}_n(1/\lambda- 1/\lambda^{\prime})\rvert^2\big)}\\ 
    \mathbf{C}_{\mathrm{B}} &= -\frac{2 \pi \Sigma_B^2 \mu_{2,r} \left ( f_{0} f_{0}^{\prime}\right)^{-\gamma}}{N^2 } \sum_{n=0}^{N} \exp{\big(- 2\pi^2\Sigma_B^2 \lvert \mathbf{u} - \mathbf{u}^{\prime} + \mathbf{x}_n/\lambda^{\prime}\rvert^2\big)}\\ 
    \mathbf{C}_{\mathrm{C}} &= -\frac{2\pi \Sigma_C^2 \mu_{2,r}\left ( f_{0} f_{0}^{\prime}\right)^{-\gamma}}{N^2} \sum_{n=0}^{N} \exp{\big(-2 \pi^2 \Sigma_C^2 \lvert \mathbf{u} - \mathbf{u}^{\prime} + \mathbf{x}_{n}/\lambda\rvert^2)}\\ 
    \end{split}
    \label{eq: beam_covariance_A_B_C}
\end{equation}

Here we define:

\begin{equation}
\begin{aligned}
    \Sigma^2_A &= \frac{\sigma^{2} \sigma^{\prime 2} \sigma^{2}_d \sigma_d^{\prime2}}{\sigma^{\prime2}\sigma_d^2 \sigma_d^{\prime2} + \sigma^2 \sigma_d^2 \sigma_d^{\prime 2} + \sigma^2 \sigma^{\prime 2}\sigma_d^2 + \sigma^2 \sigma^{\prime 2}\sigma_d^{\prime 2}}\\
    \Sigma^2_B &= \frac{\sigma^2 \sigma^{\prime 2}\sigma_d^{\prime 2}}{2\sigma^{\prime 2} \sigma_d^{\prime 2} + \sigma^{2} \sigma^{\prime 2}_d + \sigma^2 \sigma^{\prime 2}}\\
    \Sigma^2_C &= \frac{\sigma^{2} \sigma^{\prime 2} \sigma_d^{2}}
    {\sigma^{\prime 2} \sigma_d^{2} + 2 \sigma^{2} \sigma_d^{2} + \sigma^{2} \sigma^{\prime 2}}\\
\end{aligned}
\end{equation}

The 4th and fifth integral require us keep the primed and unprimed coordinates separate, we then solve the integrals separately using the same procedure. Rewrite all beams that observe the same sky, either  $l$ or $l^{\prime}$  into a single Gaussian and perform two Hankel transforms one over the unprimed and one over the primed coordinates..

\begin{equation}
    \begin{aligned}
    \mathbf{C}_{\mathrm{D}} &= (\mu_{1, \mathrm{m}} + \mu_{1, \mathrm{r}})^2 \left ( f_{0} f_{0}^{\prime}\right)^{-\gamma}\frac{2\pi \Sigma_D^2\Sigma_D^{\prime 2} }{N^3 }  \\
    & \times  \sum_{n=0}^{N} \exp{\big(-2\pi^2\Sigma_D^2 \lvert \mathbf{u} -\mathbf{x}_n/\lambda\rvert^2)} \exp{\big(-2\pi^2\Sigma_D^{\prime 2} \lvert \mathbf{u}^{\prime} - \mathbf{x}_n/\lambda^{\prime} \rvert^2)}\\ 
    \mathbf{C}_{\mathrm{E}} &=(\mu_{1, \mathrm{m}} + \mu_{1, \mathrm{r}})^2\left ( f_{0} f_{0}^{\prime}\right)^{-\gamma} \frac{2\pi^2 \Sigma_D \Sigma_D^{\prime} }{N^4} \\
    &\qquad \qquad\qquad \times\Big( \sum_{n=0}^{N} \exp{\big(-2\pi^2\Sigma_D^2 \lvert \mathbf{u} - \mathbf{x}_n/\lambda\rvert^2)} \Big)\\
    &\qquad\qquad\qquad \times \Big(\sum_{n=0}^{N} \exp{\big(-2\pi^2 \Sigma_D^{\prime 2} \lvert \mathbf{u}^{\prime} - \mathbf{x}_n/\lambda^{\prime} \rvert^2)}\Big)\\ 
    \end{aligned}
    \label{eq: beam_covariance_D_E}
\end{equation}

Where 
\begin{equation}
    \Sigma^2_D = \frac{\sigma^2 \sigma_d^2}{\sigma^2 + \sigma_d^2}
\end{equation}

These 6 terms together, effectively averaging the error over 16 dipoles form the covariance for a single baseline between different frequencies. It quantifies the error and how this is error is correlated.

\section{Linearised Gain Error Covariance Matrix}
\label{apx: linearised_gain_error_covariance}
Starting from the definition of covariance we compute $\langle \mathbf{\hat{r}}\mathbf{\hat{r}}^{\prime} \rangle$ and $\langle \mathbf{\hat{r}}\rangle \langle \mathbf{\hat{r}}^{\prime} \rangle$ to form the residual covariance matrix. Taking Equation~(\ref{eq: estimated_residuals_expanded} we can work out the averaged product assuming the gain error $\delta g$ is independent from both $\mathbf{m}$ and $\mathbf{r}$ within the same $u$-bin. This is strictly not true, but because the gain error is a linear combination of several independent $u$-bins this is a reasonable approximation.

\begin{equation}
    \begin{aligned}
    \langle \hat{r} \hat{r}^{\prime} \rangle &= \langle (\delta g + \delta g^{*})(\delta g^{\prime} + \delta g^{*\prime})\rangle \langle m m^{*\prime} \rangle\\ 
    &\qquad-\langle (\delta g + \delta g^{*})(1 - \delta g^{\prime} - \delta g^{*\prime})\rangle \langle m\rangle \langle r^{*\prime} \rangle \\
    &\qquad-\langle (1 - \delta g - \delta g^{*})(\delta g^{\prime} + \delta g^{*\prime})\rangle \langle m^{*\prime}\rangle \langle m \rangle \\
    &\qquad+\langle (1 - \delta g - \delta g^{*})(1 - \delta g^{\prime} - \delta g^{*\prime})\rangle \langle  r r^{*\prime} \rangle \\
    \end{aligned}
    \label{eq: residual_product}
\end{equation}
The product of averages is fairly straight forward to compute, when we combine both to get the covariance we notice that many terms will drop out leaving us with

\begin{equation}
    \begin{aligned}
    \mathbf{C}_{\mathrm{\hat{r}}} &= \langle (\delta g + \delta g^{*})(\delta g^{\prime} + \delta g^{*\prime})\rangle \langle m m^{*\prime} \rangle\\ 
    &\qquad - \langle (\delta g + \delta g^{*})\rangle \langle(\delta g^{\prime} + \delta g^{*\prime})\rangle \langle m\rangle \langle m^{*\prime} \rangle\\
    &\qquad+\langle (1 - \delta g - \delta g^{*})(1 - \delta g^{\prime} - \delta g^{*\prime})\rangle \langle  r r^{*\prime} \rangle \\
    &\qquad-\langle (1 - \delta g - \delta g^{*})\rangle \langle(1 - \delta g^{\prime} - \delta g^{*\prime})\rangle \langle  r\rangle \langle r^{*\prime} \rangle \\
    \end{aligned}
    \label{eq: residuall_covariance}
\end{equation}

This expression looks fairly similar to the product of gain covariance $\mathbf{C}_{g}$ with either the model covariance $\mathbf{C}_{m}$ or the residual covariance $\mathbf{C}_{r}$. Because the mean model visibilities $\langle m \rangle$ integrate to zero at $u$ scales beyond the diameter of a tile, and similarly for the residuals $\langle r \rangle$ we can all terms that contain these means. Following a similar argument we can say that $\mathbf{C}_{g} \sim \langle \delta g \delta g^{\prime} \rangle$, $\mathbf{C}_{m} \sim \langle \delta m \delta m^{\prime} \rangle$, and $\mathbf{C}_{r} \sim \langle \delta r \delta r^{\prime} \rangle$. Assuming the gain errors from two antennas are independent we can write  $2 \mathbf{C}_{g} = \langle (\delta g + \delta g^{\prime})(\delta g + \delta g^{\prime})^{*}$


\bsp	
\label{lastpage}
\end{document}